\documentclass[aps,
reprint,
superscriptaddress,
prb]{revtex4-2}

\usepackage{xcolor}
\usepackage{markdown}
\usepackage{algorithmicx}
\usepackage{booktabs}
\usepackage{algpseudocode}

\usepackage{hyperref}

\renewcommand{\selectlanguage}[1]{}

\newcommand{\bra}[1]{\langle#1|}
\newcommand{\ket}[1]{|#1\rangle}

\newcommand{\h}[0]{\hbar}
\newcommand{\tr}[0]{\mathrm{Tr}}

\newcommand{\id}[0]{\mathbb{I}}

\newcommand{\renyi}{R\'enyi-$2$}
\newcommand{\rlite}{CoMPuTE}

\newcommand{\Cnl}{\mathcal{C}_{n}^{\ell}}
\newcommand{\Cbar}[1]{\overline{#1}}
\newcommand{\rhonl}{\rho_{n}^{\ell}}

\newcommand{\lstar}{\ell^{\ast}}

\renewcommand{\textproc}[1]{\textsc{#1}}

\renewcommand{\eqref}[1]{Eq.~(\ref{#1})}
\newcommand{\xeqref}[1]{(\ref{#1})}

\begin{document}


\title{Compressed minimum-purity time evolution for late-time quantum dynamics}

\author{Moksh Bhateja}
\email[]{moksh@pks.mpg.de}
\affiliation{Max Planck Institute for the Physics of Complex Systems (MPI-PKS), 01187 Dresden, Germany}
\affiliation{Indian Institute of Space Science and Technology, Thiruvananthapuram 695547, India}
\affiliation{Forschungszentrum J\"{u}lich GmbH, Peter Gr\"{u}nberg Institute,
	Quantum Control, 52425 J\"{u}lich, Germany}
\author{Jonas B. Rigo}
\email[]{Jonas.Rigo@physik.uni-regensburg.de}
\affiliation{University of Regensburg, D-93053 Regensburg, Germany}
\affiliation{Forschungszentrum J\"{u}lich GmbH, Peter Gr\"{u}nberg Institute,
	Quantum Control, 52425 J\"{u}lich, Germany}
\author{Markus Schmitt}
\email[]{markus.schmitt@ur.de}
\affiliation{Forschungszentrum J\"{u}lich GmbH, Peter Gr\"{u}nberg Institute,
	Quantum Control, 52425 J\"{u}lich, Germany}
\affiliation{University of Regensburg, D-93053 Regensburg, Germany}

\date{\today}

\begin{abstract}
Unitary time evolution of initially simple quantum many-body states rapidly generates entanglement and complex correlations, which limits direct numerical simulations.
The late-time dynamics of physical observables, however, typically exhibits an effective simplicity in the form of hydrodynamics or kinetic theory.
This leads to the question whether microscopic equations of motion can remain accurate and tractable up to long time scales by discarding irrelevant information in a controlled manner.
Here, we introduce compressed minimum-purity time evolution (\rlite{}) as an approach to keep track of a consistent set of reduced local density matrices, closing the hierarchical equations of motion using a minimum-purity principle.
%
%
In benchmark applications we demonstrate (i) accurate description of energy diffusion in the one-dimensional mixed-field Ising model, (ii) the applicability to genuinely out-of-equilibrium Floquet dynamics starting from a pure state, and (iii) the limitations of the local reduced density matrix approximation when describing transport in the XXZ chain at $\Delta=1$ that is governed by increasingly non-local integrals of motion.
%
The \rlite{} method enhances computational efficiency in comparison to the closely related local-information time evolution algorithm \cite{kleinkvorningTimeevolutionLocalInformation2022,artiacoEfficientLargeScaleManyBody2024}, opening a possible route towards an extension to systems in higher spatial dimensions.
\end{abstract}

\maketitle

\section{Introduction}
Understanding dynamical properties is essential for an encompassing characterization of quantum many body systems.
Near equilibrium, dynamical response functions reveal essential properties like the existence of quasiparticles \cite{amentResonantInelasticXray2011}, fractionalization \cite{hanFractionalizedExcitationsSpinliquid2012}, quantum criticality \cite{gegenwartQuantumCriticalityHeavyfermion2008a} or transport behavior \cite{kjaergaardQuantizedConductanceDoubling2016}.
Far from equilibrium, identifying the principles that govern unitary many body dynamics and eventual equilibration is an active area of research \cite{polkovnikovColloquiumNonequilibriumDynamics2011,dalessioQuantumChaosEigenstate2016,mitraQuantumQuenchDynamics2018}.
In many settings the intricate long-time regime is of particular interest because it determines whether and how a system equilibrates \cite{rigolThermalizationItsMechanism2008,deutschEigenstateThermalizationHypothesis2018}.

While several experimental techniques naturally probe slow modes and late-time dynamics \cite{scheieDetectionKardarParisi2021, Wienand2024}, extending the time scales reachable by numerical methods remains a major challenge.
This difficulty may appear somewhat paradoxical because late-time behavior often admits effective descriptions for the physical observables of interest, such as hydrodynamics \cite{luxHydrodynamicLongtimeTails2014a, Wienand2024} and kinetic theory \cite{gopalakrishnanKineticTheorySpin2019a}, that are classically tractable.
However, connecting such descriptions to first-principles models requires following the dynamics across all time scales, and the build-up of complex correlations at short and intermediate times typically constitutes the central bottleneck.
The effective simplicity of late-time descriptions therefore suggests that much of the detailed high-order correlations generated at intermediate times is ultimately irrelevant for the observables of interest \cite{Dubail2017,Wang2019,Noh2020,Schmitt2022,Reid2021,vonKeyserlingk2022}.
This raises the question whether one can overcome the intermediate-time complexity bottleneck by systematically focusing on the part of the information that determines the late-time dynamics.

Equations of motion based on Bogoliubov–Born–Green–Kirkwood–Yvon (BBGKY) type hierarchies \cite{bogoliubov1947kinetic,KirkwoodBBGKY} provide a long-established class of approaches in this spirit, but they are often limited by ad hoc truncations at a given order.
More recently, a variety of methods have aimed to identify irrelevant information in a more targeted manner using controlled compression principles and information-theoretic concepts, including dissipation-assisted operator evolution \cite{Pollmann2022,vonKeyserlingk2022}, density matrix truncation \cite{whiteQuantumDynamicsThermalizing2018}, locally relaxed evolution schemes \cite{frias-perezConvertingLongRangeEntanglement2024}, Krylov recursion approaches \cite{Parker_2019,wang2023diffusion}, and local-information time evolution (LITE) \cite{kleinkvorningTimeevolutionLocalInformation2022,artiacoEfficientLargeScaleManyBody2024}. These methods provide a bridge between reversible unitary quantum dynamics and the emergent irreversibility described by statistical mechanics. A direct application is the computation of near-equilibrium transport coefficients in the late-time regime beyond the entanglement barrier.

In this work we build on LITE, a BBGKY-inspired method that evolves a consistent hierarchy of reduced density matrices up to a maximal subsystem size rather than the full density matrix. Because this finite hierarchy is not closed, LITE reconstructs missing larger subsystems from overlapping lower-level reduced density matrices (marginals) through maximum-entropy recovery and removes excess large-scale correlations by an entropy-maximizing correction that preserves the marginals and boundary currents, enabling accurate late-time transport \cite{artiacoEfficientLargeScaleManyBody2024,harkinsNanoscaleEngineeringDynamic2025}. 
Basing the algorithm on the principles of entropy maximization and information flows is, however, a choice.
In this work, we propose \rlite{}---an alternative approach employing a minimum-purity formulation of the required recovery and removal steps and using purity currents within the hierarchical structure.
On the technical level, \rlite{} replaces the von~Neumann entropy used for information quantification within LITE with R\'enyi-2 entropy as a measure of purity.
Thereby, matrix logarithms and spectral decompositions---constituting the major computational bottleneck of LITE---are avoided, reducing the computational complexity by a factor of the maximal subsystem Hilbert space dimension.

The original LITE formulation also gives access to scale-resolved diagnostics through the \emph{information lattice} introduced in Refs.~\cite{kleinkvorningTimeevolutionLocalInformation2022,artiacoEfficientLargeScaleManyBody2024}.
The \emph{local information} resolved on the information lattice is the information that is present in a parent subsystem at larger length scale but not already accessible from its overlapping marginals.
There is no direct R\'enyi-2 purity analogue of this quantity, because the induced entropy lacks a general strong-subadditivity property \cite{lindenStructureRenyiEntropic2013}.
We therefore introduce the \emph{purity gain}, the (negative or positive) gain in R\'enyi-2 purity information obtained when the overlapping marginals are combined into their parent subsystem.
Although it is not as strict as the von~Neumann irreducible information, the purity gain still provides a useful scale-resolved diagnostic.

In the following we first certify the accuracy of \rlite{} by confirming agreement of the resulting energy diffusion coefficient in a chaotic one-dimensional quantum Ising model with results obtained by alternative methods \cite{Pollmann2022,thomas2023comparing,wang2023diffusion,Parker_2019}.
Furthermore, we address two demanding cases where the extensions of \rlite{} are decisive, namely driven Floquet dynamics with a pure initial state and transport in the integrable XXZ chain.
In the Floquet setting, \rlite{} reproduces the relaxation and heating trends obtained with density matrix truncation \cite{whiteQuantumDynamicsThermalizing2018} over the full accessible time window while retaining access to purity diagnostics.
The dynamics of the XXZ chain puts the underlying principles of the LITE algorithm to the test, illustrating the limitations of the subsystem approach to state compression:
While energy transport is well captured up to the latest times, the duration of the superdiffusive spin transport observed in the simulation is bounded by the truncation scale because the relevant contributions rely on increasingly nonlocal operators.
At the same time, the improved scaling of \rlite{} extends the accessible time window and makes the intermediate-time transport regime visible and systematically improvable by increasing the maximum subsystem size.

The paper is organized as follows.
Section~\ref{sec:background} reviews the reduced-density-matrix hierarchy behind LITE and introduces purity gain and purity currents.
Section~\ref{sec:renyi2} derives the minimum purity-based recovery and removal rules.
Section~\ref{sec:results} presents benchmarks for diffusive transport, driven dynamics, and integrable transport.

\begin{figure}[t]
    \centering
    \includegraphics[width=1.\linewidth]{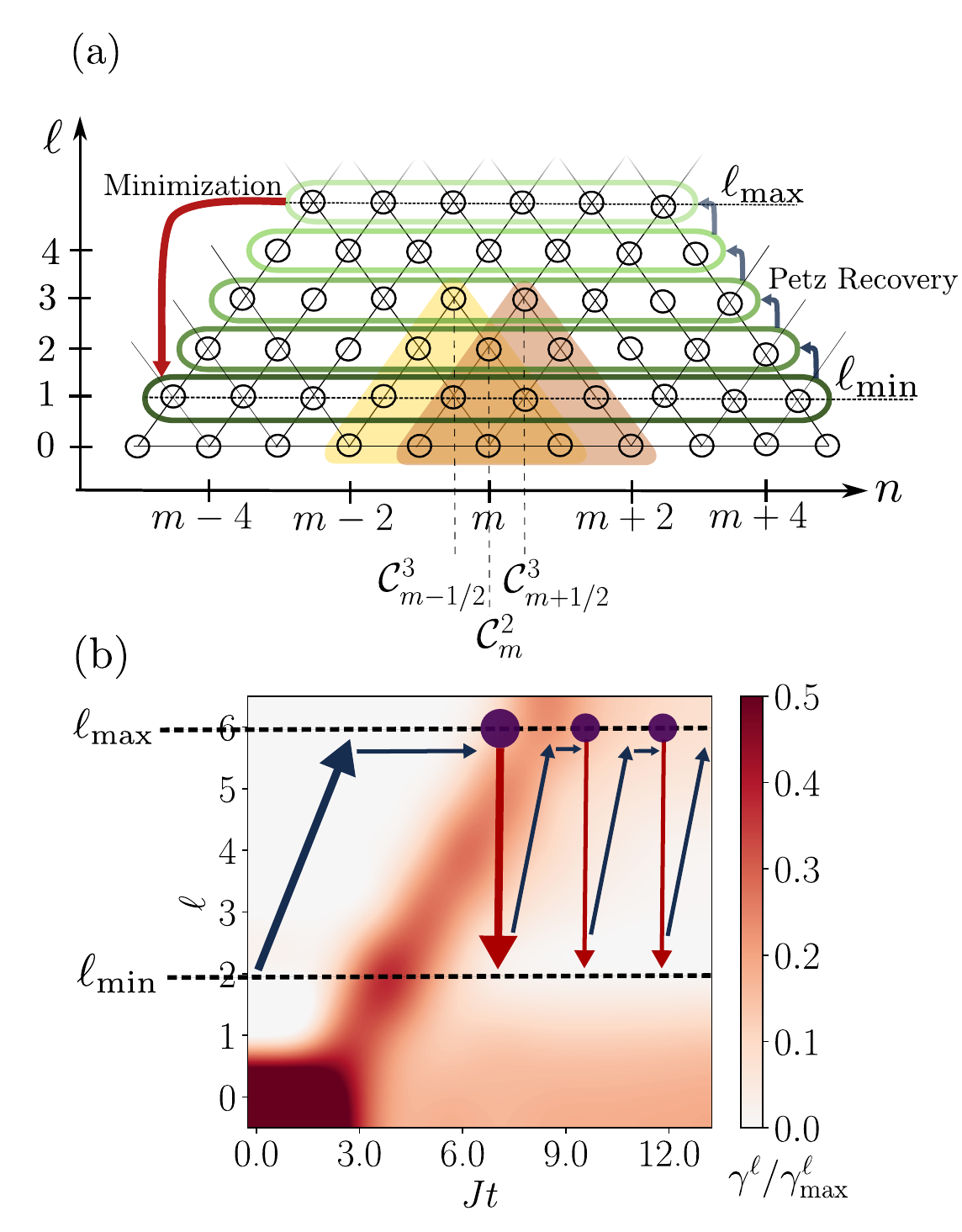}
    \caption{
    Schematic illustration of the \rlite{} control cycle on the subsystem hierarchy, analogous to the LITE approach.
    (a) Reduced density matrices $\rho_n^\ell$ are retained up to the maximum level $\ell_{\max}$.
    Higher-level density matrices required by the hierarchical equations of motion are reconstructed from overlapping lower-level marginals by recovery, as indicated by the shaded regions.
    Once the top retained level has accumulated too much large-scale structure, a removal step resets the active evolution to the lower truncation level $\ell_{\min}$.
    (b) Purity gain for Hamiltonian Eq.~\ref{eq:mixed_ising} (\(h_L = 0.25J,~h_T= -0.525J\)) and initial state \(\rho_{\rm init} = \bigotimes^7_{n=1}\big( \frac{2}{3} \ket{\uparrow}\bra{\uparrow} + \frac{1}{3}\ket{\downarrow}\bra{\downarrow} \big) \)) shown to move through levels during time evolution.
    The alternation between upward purity-gain flow, recovery near \(\ell_{\max}\), and removal back to \(\ell_{\min}\) forms the basic control cycle used by \rlite{}.
    }
    \label{fig:LITE_recap}
\end{figure}

\section{Subsystem lattice and hierarchical equations of motion}
\label{sec:background}

This section reviews the elements of LITE and introduces the new concepts needed to formulate \rlite{}.
The algorithm evolves local reduced density matrices organized by position and subsystem size, leading to an open hierarchy of equations of motion whenever the maximum retained subsystem size is smaller than the full system.
The closure combines maximum-entropy recovery of missing larger subsystems with entropy maximization on the largest retained subsystems, subject to additional constraints that preserve the current of the chosen local information measure.
In the original implementation this measure is derived from the von~Neumann entropy, whereas in \rlite{} it is derived from the R\'enyi-2 entropy, or equivalently from the purity.

\subsection{Subsystem decomposition and purity gain}
\label{subsec:info_lattice}
LITE compresses the full density matrix by representing the system state as a collection of reduced density matrices up to a maximum subsystem size.
This collection must be consistent, which implies that whenever two retained subsystems overlap, tracing either density matrix down to the common region must give the same reduced state.
We introduce this construction for a one-dimensional spin-$1/2$ chain of $N$ sites with local Hilbert-space dimension $d=2$, but it can be generalized to any other one-dimensional system.

A contiguous subsystem of $\ell+1$ sites centered around $n$ is denoted by $\Cnl$,
\begin{equation}
\Cnl \equiv \{n-\ell/2, n-\ell/2+1, \dots, n+\ell/2\}.
\end{equation}
The integer $\ell=0,1,\dots,N-1$ labels the subsystem level and $\ell+1$ is the subsystem size.
When $\ell$ is even, $n$ is an integer and when $\ell$ is odd, $n$ is a half-integer.
Given a global density matrix $\rho$, the reduced state on $\Cnl$ is
\begin{equation}
\rhonl := \tr_{\Cbar{\Cnl}}(\rho),
\label{eq:red_dm}
\end{equation}
where $\Cbar{\Cnl}$ denotes the complement of $\Cnl$.
The set of all $\Cnl$ forms a subsystem hierarchy because each node is built from two overlapping nodes at the previous level,
\begin{equation}
\Cnl = \mathcal{C}_{n-\frac12}^{\ell-1} \cup \mathcal{C}_{n+\frac12}^{\ell-1},
\end{equation}
which creates the \textit{subsystem lattice} shown in Fig.~\ref{fig:LITE_recap} (a), whose nodes $(n,\ell)$ are the subsystems $\Cnl$.
Using the subsystem-lattice construction, LITE compresses the full density matrix by introducing a \textit{working level} $\lstar$ and representing the state by the collection $\{\rho_n^{\lstar}\}_n$ of reduced density matrices on all retained subsystems of size $\lstar+1$.

For any entropy functional $S$, we define the information at node $(n,\ell)$
\begin{equation}
I_n^{\ell} := \ln\!\big(d^{\ell+1}\big) - S(\rhonl),
\label{eq:Inl_def}
\end{equation}
which vanishes for a maximally mixed subsystem and grows as the reduced state becomes more structured.
For \rlite{}, the R\'enyi-2 entropy and purity are, respectively,
\begin{equation}
S_2(\rho)=-\log\tr(\rho^2),
\qquad
P(\rho)=\tr(\rho^2).
\end{equation}
For this choice, the corresponding node information $I_n^{\ell}$ measures the local purity content relative to the maximally mixed state.

To isolate the information associated with a particular scale, we account for the overlap in the hierarchical construction. The two level-$(\ell-1)$ children of $\mathcal{C}_n^\ell$ share the level-$(\ell-2)$ subsystem $\mathcal{C}_n^{\ell-2}$. The scale- and site-resolved contribution is therefore obtained from the overlap-corrected combination
\begin{equation}
\gamma_n^{\ell}
:=
I_n^{\ell}
-
I_{n-\frac12}^{\ell-1}
-
I_{n+\frac12}^{\ell-1}
+
I_n^{\ell-2},
\label{eq:mutual_info}
\end{equation}
which gives the scale-resolved contribution after summing over all sites,
\begin{equation}
    \gamma^\ell = \sum_n \gamma_n^\ell.
\end{equation}
For the R\'enyi-2 choice, we call \(\gamma_n^{\ell}\) the \emph{log purity gain} at scale \(\ell\) and position \(n\). In the following, we drop ``log'' and refer to \(\gamma_n^{\ell}\) simply as the \emph{purity gain}. This refers to the overlap-corrected gain in local purity information when passing from the two lower-level children to their parent subsystem. For the tripartition
\(ABC=\Cnl\),
\(AB=\mathcal{C}_{n-\frac12}^{\ell-1}\),
\(BC=\mathcal{C}_{n+\frac12}^{\ell-1}\), and
\(B=\mathcal{C}_{n}^{\ell-2}\), this gives
\begin{align}
    \label{eq:renyi2_signed_info}
    \gamma_n^{\ell}
    &=
    S_2(AB)+S_2(BC)-S_2(B)-S_2(ABC) \\
    &=
    \log\frac{P_{ABC}P_B}{P_{AB}P_{BC}},
\end{align}
where \(P_X=\tr(\rho_X^2)\). Thus \(\gamma_n^{\ell}\) compares the purity information of the parent subsystem with the information already contained in its two overlapping children, with the shared overlap \(B\) added back once. 

For a pure state \(\rho_{ABC} \equiv \ket{\psi}\bra{\psi}\), the top-level purity is always \(P(\rho_{ABC}) = 1\), whereas the marginals are generally mixed, \(P(\rho_{AB}),~P(\rho_{BC}) \le 1\). This implies that purity is gained when going from smaller to larger subsystems, resulting in a positive purity gain characterized by
\[
P_{ABC}P_B>P_{AB}P_{BC}.
\]
In general, however, the purity gain can also be negative when
\[
P_{ABC}P_B<P_{AB}P_{BC},
\]
which reflects the fact that the R\'enyi-2 entropy does not satisfy a general strong-subadditivity theorem. Consequently, the overlap-corrected R\'enyi-2 quantity \(\gamma_n^{\ell}\) is signed, whereas the corresponding von~Neumann irreducible information is non-negative by strong subadditivity \cite{lindenStructureRenyiEntropic2013}.

Importantly, the next section shows that the local R\'enyi-2 entropy, or equivalently the local purity information, changes only through boundary terms in the subsystem equation of motion. The purity gain therefore inherits a local continuity structure. It changes through currents between adjacent nodes on the subsystem lattice and can be viewed as a diagnostic of how local purity information is redistributed over position and scale during the evolution.

\subsection{Subsystem evolution and hierarchy closure}
\label{subsec:lite_recap}

We now summarize how LITE turns the subsystem hierarchy into a finite time-evolution scheme.
The starting point is the von~Neumann equation for a density matrix $\rho$ and a Hamiltonian $H$ with $\h=1$,
\begin{equation}
\label{eq:von_neumann}
\partial_t \rho = -i[H,\rho].
\end{equation}
The discussion below is written for nearest-neighbor interactions.
Longer-ranged local Hamiltonians can be handled by the same construction \cite{artiacoEfficientLargeScaleManyBody2024}.

\subsubsection{Hierarchical equation of motion for local density matrices}
Tracing out the complement of $\mathcal{C}_n^\ell$ transforms \eqref{eq:von_neumann} into an exact hierarchical (BBGKY-like) equation of motion for the $\rho_n^\ell$ defined in \eqref{eq:red_dm} of the form
\begin{widetext}
\begin{equation}
\partial_t \rho_n^\ell
=
-i[H_n^\ell,\rho_n^\ell]
-i\,\tr_{L}\!\left([H_{n-\frac{1}{2}}^{\ell+1} - H_n^\ell,\;\rho_{n-\frac{1}{2}}^{\ell+1}]\right)
-i\,\tr_{R}\!\left([H_{n+\frac{1}{2}}^{\ell+1} - H_n^\ell ,\;\rho_{n+\frac{1}{2}}^{\ell+1}]\right),
\label{eq:subsystem_eom}
\end{equation}
\end{widetext}
where $H_n^\ell$ denotes an effective subsystem Hamiltonian defined by tracing the full Hamiltonian over the complement of $\mathcal{C}_n^\ell$,
$H_n^\ell = \tr_{\Cbar{\Cnl}}(H)/\dim(\bar{\mathcal{C}}_n^\ell)$, 
and $\tr_L$ ($\tr_R$) traces out the leftmost (rightmost) site of $\mathcal{C}_n^{\ell+1}$.
In the second and third terms, $H_n^\ell$ is understood as embedded into the $(\ell+2)$-site spaces as $\id \otimes H_n^\ell$ and $H_n^\ell\otimes \id_{0}$, respectively.
\eqref{eq:subsystem_eom} is clearly not an independent equation for all $\rho^\ell_n$, since evolving level $\ell$ requires access to density matrices at level $\ell+1$, generating a hierarchy that only closes at the top level (the full system) \cite{artiacoEfficientLargeScaleManyBody2024}.
The apparent locality in \eqref{eq:subsystem_eom}, which couples the subsystem $\rho_n^\ell$ on subsystem-lattice node $(n,\ell)$ to the subsystems on nodes $(n-\frac{1}{2},\ell+1)$ and $(n+\frac{1}{2},\ell+1)$, is a direct consequence of Hamiltonian locality since only terms acting across the left or right boundary of $\mathcal{C}_n^\ell$ can change its reduced state, and these terms are contained in the two neighboring enlarged subsystems.

\eqref{eq:subsystem_eom} not only determines the dynamics of the reduced density matrices, it also induces a local flow of R\'enyi-2 entropy on the subsystem lattice. This flow satisfies a continuity equation because it inherits the locality of \eqref{eq:subsystem_eom}, making R\'enyi-2 entropy a locally transported quantity on the subsystem lattice. Under exact unitary evolution the global purity \(P(\rho)=\tr(\rho^2)\) is conserved. 
By contrast, the local purity \(P_n^\ell=\tr[(\rho_n^\ell)^2]\) of a reduced density matrix changes through Hamiltonian terms coupling \(\mathcal{C}_n^\ell\) to its neighboring sites. The internal Hamiltonian contribution drops out by cyclicity of the trace, so the time derivative of \(S_2(\rho_n^\ell)\) can be written entirely in terms of boundary currents from the two neighboring level-\(\ell+1\) subsystems. These currents are 
\begin{align}
    J_L(\rho_n^{\ell}) 
    &= -i\tr\Big(\big[ \nabla_\rho S^\ell_n,
    H^{\ell+1}_{n - 1/2} - H^\ell_n \big]\rho^{\ell+1}_{n-1/2}\Big),
    \label{eq:renyi_currents_intro_L}\\
    J_R(\rho_n^{\ell}) 
    &= -i\tr\Big(\big[ \nabla_\rho S^\ell_n,
    H^{\ell+1}_{n + 1/2} - H^\ell_n \big]\rho^{\ell+1}_{n+1/2}\Big),
    \label{eq:renyi_currents_intro_R}
\end{align}
with
\begin{equation}
    \nabla_\rho S^\ell_n
    =
    - \frac{2}{\tr[(\rho^\ell_n)^2]}
    \big(\rho^\ell_n\big)^T.
\end{equation}
The corresponding currents of local purity information \(I=\log d-S_2\) have the opposite sign.

\subsubsection{Truncating the hierarchy}

In practice, the evolution starts at the minimal retained level $\ell_{\min}$ and proceeds at a working level $\lstar$ within the scale window $\ell_{\min} \le \lstar \le \ell_{\max}$ shown in Fig.~\ref{fig:LITE_recap} (a). Thus $\ell_{\min}+1$ is the smallest subsystem size kept explicitly, while $\ell_{\max}+1$ is the largest subsystem size allowed before truncation.
The equation of motion \eqref{eq:subsystem_eom} also requires higher-level states $\{\rho^{\lstar+1}_{n\pm 1/2}\}$.
Whenever these states are not stored explicitly, LITE closes the hierarchy by a \textit{recovery step} based on the \textit{projected Petz recovery map} \cite{petzSufficientSubalgebrasRelative1986a,kleinkvorningTimeevolutionLocalInformation2022,artiacoEfficientLargeScaleManyBody2024}. This step reconstructs a higher-level density matrix from two overlapping reduced subsystems as the maximum-entropy state whose left and right partial traces reproduce the input subsystems.
For a reduced subsystem at level $\lstar+1$ centered at $n+1/2$, recovery constructs a tripartite state $\rho_{ABC}\equiv \rho^{\lstar+1}_{n+1/2}$ from two overlapping marginals $\rho_{AB}\equiv \rho^{\lstar}_{n}$ and $\rho_{BC}\equiv \rho^{\lstar}_{n+1}$, consistent with the subsystem-lattice relation $\mathcal{C}^{\lstar+1}_{n+1/2}=\mathcal{C}^{\lstar}_{n}\cup \mathcal{C}^{\lstar}_{n+1}$.
The key to maintaining a compressed representation of the system is to avoid recovering any subsystem with scale beyond $\ell_{\max}$.
Once the working level reaches $\lstar=\ell_{\max}$, accumulated top-level purity gain can trigger a reset back to $\ell_{\min}$, as illustrated in Fig.~\ref{fig:LITE_recap}(a). The criterion for this reset is described in the following.

As the dynamics proceeds, purity gain tends to move toward larger subsystem sizes, as shown in Fig.~\ref{fig:LITE_recap}(b).
The algorithm monitors the positive part of $\gamma_n^{\lstar}$
\begin{equation}
    \gamma_{n,\mathrm{pos}}^{\ell}=\max\!\left(\gamma_n^{\ell},0\right)
\end{equation}
and promotes $\lstar\to\lstar+1$ whenever it exceeds a small threshold $q_{\lstar}$ at any position.
The negative purity gains are clipped to zero because that prevents negative purity gains from canceling positive buildup at the top level $\ell_{\max}$ in the truncation criteria.
In \rlite{} we therefore keep the signed value
\begin{equation}
    \gamma_{n,\mathrm{signed}}^{\ell}=\gamma_n^{\ell}
\end{equation}
for diagnostics and conservation checks, but use the clipped positive part
\begin{equation}
    \gamma_{n,\mathrm{pos}}^{\ell}=\max\!\left(\gamma_n^{\ell},0\right)
\end{equation}
for algorithmic control thresholds.
The full purity gain $\gamma_{n}^{\ell}$ records cases in which the R\'enyi-2 overlap correction overcompensates and is used only for diagnostics.

Once $\lstar$ reaches $\ell_{\max}$, the retained hierarchy can no longer be extended.
When the positive top-level purity gain $\gamma_{\mathrm{pos}}^{\ell_{\max}} = \sum_n \gamma_{n,\mathrm{pos}}^{\ell_{\max}}$ becomes too large relative to the total stored positive gain $\gamma_{\mathrm{pos,tot}} = \sum_{n,\ell} \gamma_{n,\mathrm{pos}}^{\ell}$, \rlite{} performs current-constrained \emph{entropy maximization}, equivalently \emph{information minimization}.
This step demotes the working level back to $\ell_{\min}$ by partial tracing and then updates the working-level states by adding an allowed correction $\chi^{\ell_{\min}}_n$.
The correction maximizes the R\'enyi-2 entropy, equivalently minimizes the purity, while preserving the left and right marginals and the purity currents across the boundary $\ell_{\min}-1 \leftrightarrow \ell_{\min}$.
The reset removes excess large-scale purity structure while keeping local correlations and boundary currents intact.
Time evolution then resumes with $\lstar = \ell_{\min}$, which closes the loop and leads to the promotion-demotion cycle illustrated in Fig.~\ref{fig:LITE_recap}(b).

The two truncation scales $\ell_{\min}$ and $\ell_{\max}$ therefore play complementary roles.
The upper level $\ell_{\max}$ sets the largest explicitly represented subsystem, of size $\ell_{\max}+1$, and determines how far purity information can move to larger scales before the hierarchy has to be truncated.
It also controls the accuracy of recovery near the top of the retained hierarchy.
The lower level $\ell_{\min}$ sets the scale to which the hierarchy is reset and on which the removal correction is applied.
It therefore determines how directly the reset feeds back into the subsequent dynamics.
Empirically we have found that, once $\ell_{\max}$ is large enough, transport observables are usually more sensitive to $\ell_{\min}$ and to the removal threshold $q_{\max}$.
This motivates taking $\ell_{\max}$ as large as feasible and using $\ell_{\min}$ and $q_{\max}$ as the primary convergence parameters.

\subsubsection{Initial conditions and simulations in the thermodynamic limit}
The compression scheme employed by LITE limits the correlations and structure of usable initial states.
A simple class of allowed initial states is given by finite product states,
\begin{equation}
\rho_{\mathrm{init}}=\bigotimes_{m=1}^{N}\rho_m ,
\end{equation}
where $\rho_m$ can be pure $\rho_m = \ket{\psi_m}\bra{\psi_m}$ or mixed $P(\rho_m) < 1$ within \rlite{}.
For such states every reduced density matrix is obtained by tensoring the corresponding one-site density matrices.
The initial hierarchy can therefore be constructed without storing long-range correlations.
This makes finite product states a natural starting point for finite-size simulations and for pure-state protocols such as the driven example discussed below.

For transport problems, a localized perturbation of an infinite-temperature background is sufficient to probe the diffusion behavior of the system.
This choice combines a nontrivial local initial condition with an environment whose reduced density matrices are maximally mixed and carry no local purity content.
For an initial local energy hot spot $\rho_{n,\mathrm{init}}^\ell$ embedded in such a background,
\begin{equation}
\rho_{\mathrm{init}}
=
\Big(\bigotimes_{m<n-\ell/2}\rho_{m,\infty}\Big)
\otimes \rho_{n,\mathrm{init}}^\ell
\otimes
\Big(\bigotimes_{m>n+\ell/2}\rho_{m,\infty}\Big),
\label{eq:gen_init_state}
\end{equation}
with $\rho_{m,\infty}$ maximally mixed, the nontrivial dynamics is initially confined to a finite region in real space.
The infinite-temperature environment can then be represented by a finite simulation window surrounding the hot spot.
As time evolves, structure spreads outward and eventually approaches the boundary of this effective system.
A padding step adds or removes boundary sites represented by maximally mixed one-site density matrices.
Padding is triggered when the purity gain near a boundary exceeds a small threshold $p$, which keeps the evolving dynamics separated from the effective system boundaries.

\section{Reducing computational complexity with R\'enyi-2 entropy}
\label{sec:renyi2}

From a computational perspective, the time complexity of the LITE algorithm is bounded by the method used to compute the entropy in \eqref{eq:Inl_def} and by the core subroutines used for information removal and the Petz recovery map.
If the von~Neumann entropy is chosen to measure information content, as in Refs.~\cite{artiacoEfficientLargeScaleManyBody2024,kleinkvorningTimeevolutionLocalInformation2022}, then $S(\rho)=-\tr[\rho \log\rho]$,
which requires evaluating $\log\rho$ and therefore repeated matrix decompositions, leading to a cost that scales as $\mathcal{O}(d_\ell^3)$ for a level-$\ell$ density matrix.

Here we instead adopt the \renyi{} entropy,
\begin{equation}
	S_2(\rho) \equiv -\log \tr(\rho^2),
\end{equation}
and define the local purity information $I(\Cnl)$ by the same substitution $S \mapsto S_2$ in \eqref{eq:Inl_def}.
The local purity information of $\rho^\ell_n$ can then be computed from $\tr[(\rho^\ell_n)^2]$ without explicitly forming the matrix product $(\rho^\ell_n)^2$, since for Hermitian $\rho^\ell_n$ one has $\tr[(\rho^\ell_n)^2]=\sum_{ij}|\rho^\ell_{n,ij}|^2$.
This yields a cost of $\mathcal{O}(d_\ell^2)$ for entropy evaluation instead of the $\mathcal{O}(d_\ell^3)$ cost associated with matrix logarithms in the von~Neumann entropy.

This reduction in complexity does not, however, by itself eliminate matrix decompositions in the \emph{information-removal} and \emph{recovery} subroutines. Crucially, both can be formulated as constrained optimization problems of the form
\begin{equation}
	\label{eq:general_opti_problem}
	\tilde{\sigma}
	=
	\underset{\sigma}{\arg\,\min}\ \mathcal{L}(\sigma)
	\qquad
	\textrm{such that}
	\qquad
	\mathcal{B}(\sigma)=0\;,
\end{equation}
where $\mathcal{B}$ is a linear map encoding the LITE constraints discussed in Sec.~\ref{subsec:lite_recap} and $\mathcal{L}$ is a convex function.
Since the logarithm is monotone, maximizing $S_2(\sigma)$ is equivalent to minimizing the purity,
\begin{equation}
	\arg\max_{\sigma}\, S_2(\sigma)
	=
	\arg\min_{\sigma}\, \tr(\sigma^2)\;.
	\label{eq:renyi2_purity_equiv}
\end{equation}
Under the condition that $\tr(\sigma)=1$, minimizing $\tr(\sigma^2)$ is equivalent to minimizing the Hilbert--Schmidt distance to the maximally mixed state,
\begin{equation}
	\tr(\sigma^2) - \frac{1}{d}
	=
	\tr\!\left[\left(\sigma-\frac{\id}{d}\right)^2\right]
	\equiv
	\left\|\sigma-\frac{\id}{d}\right\|_2^2,
	\label{eq:hs_projection_identity}
\end{equation}
with $d$ the dimension of the relevant Hilbert space.
Equations~(\ref{eq:renyi2_purity_equiv}) and (\ref{eq:hs_projection_identity}) recast the information minimization and Petz recovery routines as constrained least-squares problems.
In particular, for linear constraints $\mathcal{B}(\sigma)=0$, the constrained optimum is the projection of $\id/d$ onto the space of feasible matrices.

In the information-removal step, the constrained optimization variable is the correction $\chi^{\ell_{\min}}_n$ applied to the working-level state
\begin{equation}
	\rho_n^{\lstar}\;\mapsto\;\rho_n^{\lstar}+\chi_n^{\lstar} \;.
\end{equation}
The constraints enforce that the update does not change the left and right boundary marginals and preserves the information currents across the boundary $\ell_{\min}-1 \leftrightarrow \ell_{\min}$.
We write these constraints compactly as $\mathcal{B}_{\mathrm{IR}}(\chi)=0$.
With the \renyi{} choice, the update can be posed as
\begin{equation}
	\chi^{\ell_{\min}}_n
	=
	\arg\min_{\chi}\
	\left\|
	\left(\rho^{\ell_{\min}}_n+\chi\right)-\frac{\id}{d_{\ell_{\min}}} \;
	\right\|_2^2
\label{eq:IR_ls}
\end{equation}
such that $\mathcal{B}_{\mathrm{IR}}(\chi)=0$. 
Instead of solving \eqref{eq:IR_ls} directly, we approximate the solution to zeroth order.
That is, we use the unconstrained minimizer $\tilde{\chi}=\id/d_{\ell_{\min}}-\rho^{\ell_{\min}}_n$ and enforce the constraints by orthogonal projection onto the feasible linear subspace defined by $\mathcal{B}_{\mathrm{IR}}(\chi)=0$.
The corresponding kernel projector $P_{\ker \mathcal{B}_{\mathrm{IR}}}$ is the trace- and current-preserving projector constructed explicitly in Eqs.~(\ref{eq:ir_trace_projector_left})--(\ref{eq:ir_total_projector}) \cite{artiacoEfficientLargeScaleManyBody2024} and applying it gives the zeroth-order information-minimization correction
\begin{equation}
	\chi^{\ell_{\min},(0)}_n
	=
	P_{\ker \mathcal{B}_{\mathrm{IR}}}
	\!\left(\frac{\id}{d_{\ell_{\min}}}-\rho^{\ell_{\min}}_n\right).
	\label{eq:IR_zeroth}
\end{equation}
This is ``zeroth order'' in the sense that it applies a single closed-form projection step rather than an iterative constrained minimization.
In practice, it removes the component of $\id/d_{\ell_{\min}}-\rho^{\ell_{\min}}_n$ that would violate the boundary-marginal and current-preservation conditions, while discarding the remaining information at the working level as much as permitted by the constraints.

In the recovery step, the optimization variable is the reconstructed tripartite state $\rho_{ABC}$, constrained by the two overlapping marginals
\begin{equation}
	\tr_C(\rho_{ABC})=\rho_{AB},
	\qquad
	\tr_A(\rho_{ABC})=\rho_{BC}.
	\label{eq:marginal_constraints_intro}
\end{equation}
With the \renyi{} choice, recovery is again a constrained minimum-purity problem and we show in Appendix~\ref{app:proofs2} that it has a unique optimum
\begin{equation}
	\rho^{(0)}_{ABC}
	=
	\rho_{AB} \otimes \frac{\id_C}{d_C}
	+
	\frac{\id_A}{d_A} \otimes \rho_{BC}
	-
	\frac{\id_A}{d_A} \otimes \rho_{B} \otimes \frac{\id_C}{d_C},
	\label{eq:petz_zeroth}
\end{equation}
where $\rho_B=\tr_A(\rho_{AB})=\tr_C(\rho_{BC})$.

Although \eqref{eq:petz_zeroth} represents the true global optimum of the quadratic loss in \eqref{eq:hs_projection_identity} over the constraint set \eqref{eq:marginal_constraints_intro}, positive semidefiniteness is an additional inequality constraint that is not enforced and is therefore not guaranteed.
The same caveat applies to the projected information-removal update \eqref{eq:IR_zeroth}, which is an affine Hilbert--Schmidt projection and therefore does not, by itself, impose $\rho\succeq0$ after the update.
This is analogous to the projected Petz recovery used in the original LITE construction, where marginal constraints and current-preservation constraints are enforced at the projection level while positivity is not part of the inexpensive closed-form problem.
Imposing both \eqref{eq:marginal_constraints_intro} and $\rho_{ABC}\succeq 0$ would require a more expensive constrained optimization, while the optimum \eqref{eq:petz_zeroth} is available in closed form.
In addition, the absence of strong subadditivity for R\'enyi-2 entropy means that negative $\gamma_n^{\ell}$ can occur even for perfectly valid positive density matrices. Any negative signed purity-gain value is therefore not used as a criterion for the physicality of the density matrices.
In any case, we find empirically \eqref{eq:petz_zeroth} to be sufficiently accurate and numerically stable in the regimes studied here, and we therefore adopt it as our recovery update without a subsequent optimization under the semi-definite constraint, while monitoring physicality, marginal consistency, and the negative signed purity-information contribution described above.

\section{Results}
\label{sec:results}

In this section we benchmark our \rlite{} formalism against the original LITE formulation and demonstrate the resulting increase in accessible truncation scales. We focus first on diffusive energy transport in the mixed-field Ising chain, following the same benchmark strategy as Refs.~\cite{kleinkvorningTimeevolutionLocalInformation2022,artiacoEfficientLargeScaleManyBody2024}.
We then study a driven Floquet system and transport in the integrable XXZ spin chain, which we compare to density matrix truncation (DMT) and analytic results, respectively \cite{whiteQuantumDynamicsThermalizing2018}.

\subsection{Benchmarking on diffusive energy transport}
\label{subsec:benchmarking}

\begin{figure}[t]
    \centering
    \includegraphics[width=\linewidth]{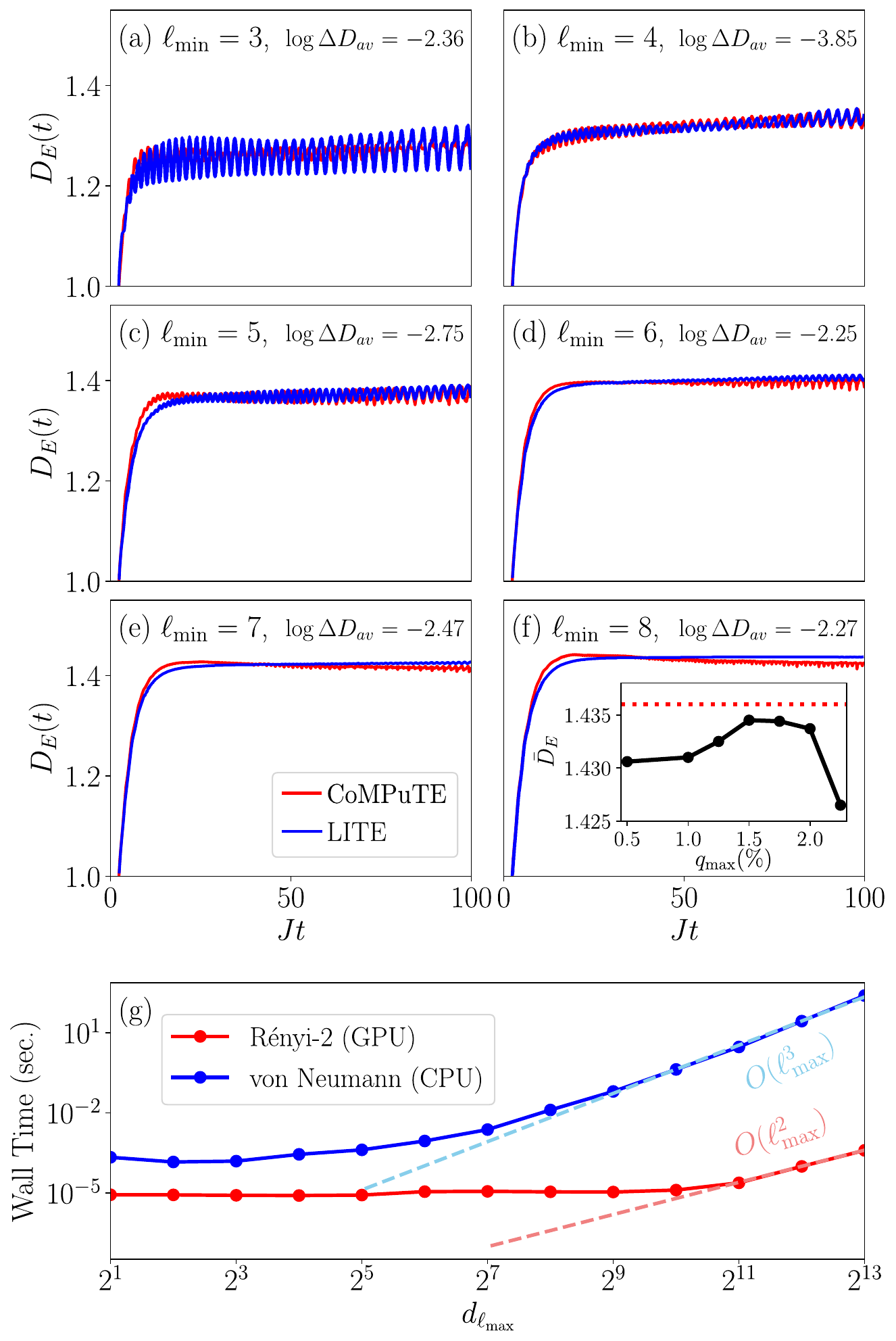}
    \caption{
        Panels (a)–(f) show the time-dependent diffusion coefficient $D_E(t)$ obtained with \rlite{} (red) and LITE (blue) for increasing working scale $\ell_{\min}=3,\dots,8$ at fixed $\ell_{\max}=9$. Each panel is computed for the information-removal threshold of $q_{\max}=0.5\%$. The annotation $\log \Delta D_{\mathrm{av}}$ reports the logarithm of the average absolute deviation between the time-averaged diffusion coefficient, quantifying that \rlite{} closely matches LITE across all shown $\ell_{\min}$. The inset in (f) shows the extracted plateau value $\bar D_E$ as a function of $q_{\max}$, demonstrating a broad stability window in which $\bar D_E$ depends only weakly on $q_{\max}$.  (g) Empirical runtime scaling of entropy evaluation versus subsystem Hilbert-space dimension $d_{\ell_{\max}}=2^{\ell_{\max}+1}$. The largest point corresponds to $d_{\ell_{\max}}\simeq 8192$, i.e.\ $(\ell_{\max}+1)=13$ spins. At this size, \rlite{} is faster by roughly six orders of magnitude and exhibits the expected weaker growth with $d_{\ell_{\max}}$.
    }
    \label{fig:benchmarking}
\end{figure}

LITE is tailored to capture \emph{local} relaxation and transport by evolving a consistent family of reduced density matrices up to a maximum subsystem size $\ell_{\max}$, and is therefore most naturally used to characterize thermalization and hydrodynamic spreading of conserved quantities rather than long-range entanglement structure. In practice, transport regimes can be distinguished by monitoring the spread of an initially localized perturbation.

To demonstrate that \rlite{} reproduces the same transport observables as LITE, we treat the same mixed-field Ising model as Ref.~\cite{artiacoEfficientLargeScaleManyBody2024}. Writing the Hamiltonian as $H=\sum_m h_m$ with two-local contributions,
\begin{equation}
\label{eq:mixed_ising}
h_m
=
J \sigma^z_m \sigma^z_{m+1}
+\frac{1}{2}\Big[
h_L(\sigma^z_m+\sigma^z_{m+1})
+h_T(\sigma^x_m+\sigma^x_{m+1})
\Big],
\end{equation}
we fix $J=1$, $h_T=1.4$, and $h_L=0.9045$, which places the system deep in the nonintegrable regime with rapid entanglement growth \cite{artiacoEfficientLargeScaleManyBody2024,kleinkvorningTimeevolutionLocalInformation2022}. As in Refs.~\cite{artiacoEfficientLargeScaleManyBody2024,kleinkvorningTimeevolutionLocalInformation2022}, we initialize an infinite-temperature environment with a small energy inhomogeneity $\rho_{n,\text{init}}^{\ell}
=
\frac{1}{Z}e^{-\beta H_n^{\ell}}$ of linear size $\ell=2$ (spanning 3 sites),
\begin{equation}
\rho_{\text{init}}
=
\Big(\bigotimes_{m<n-\ell/2}\rho_{m,\infty}\Big)
\otimes \rho_{n,\text{init}}^{\ell}
\otimes
\Big(\bigotimes_{m>n+\ell/2}\rho_{m,\infty}\Big),
\label{eq:initial_state}
\end{equation}
with $\rho_{m,\infty}$ maximally mixed, $\beta=0.005$, and $Z=\tr[e^{-\beta H_n^{\ell}}]$. As information spreads to greater scales during the time evolution, sites can be dynamically added to support the spread. Thus, we start the simulation with $N=9$ sites, and by the end of the simulation the system has reached $N=109\sim141$ (depending on the choice of $\ell_{\max}$). We evolve from $Jt=0$ to $Jt=100$.

We quantify energy transport via the second central moment of the local energy profile,
\begin{equation}
\sigma_E^2(t)=
\sum_m \big(m-\overline{m}(t)\big)^2\,\frac{E_m^r(t)}{\langle H\rangle},
\end{equation}
where $E_m^r(t)\equiv\tr\!\big(\rho(t)\,h_m^r\big)$ denotes the local energy expectation value and $\langle H\rangle=\sum_m E_m^r(t)$ is conserved in the exact dynamics. The corresponding energy center of mass is
\begin{equation}
\overline{m}(t)=\sum_m m\,\frac{E_m^r(t)}{\langle H\rangle}.
\end{equation}
From $\sigma_E^2(t)$ we define the time-dependent diffusion coefficient
\begin{equation}
D_E(t)=\frac{1}{2}\frac{d}{dt}\,\sigma_E^2(t),
\end{equation}
which typically separates a short-time transient, often close to ballistic spreading, from the late-time diffusive plateau in generic nonintegrable systems \cite{artiacoEfficientLargeScaleManyBody2024}. If a system reaches the diffusive plateau we compute the time-averaged diffusion coefficient
\begin{equation}
\bar{D}_E=\frac{1}{t_2-t_1}\int_{t_1}^{t_2}\!dt\,D_E(t)
\end{equation}
with $(t_1/J,t_2/J)=(20,100)$, which isolates the diffusive regime while averaging over residual oscillations.

\subsubsection{Accuracy benchmark: LITE vs \rlite{}.} %

\begin{figure}[t]
    \centering
    \includegraphics[width=0.95\linewidth]{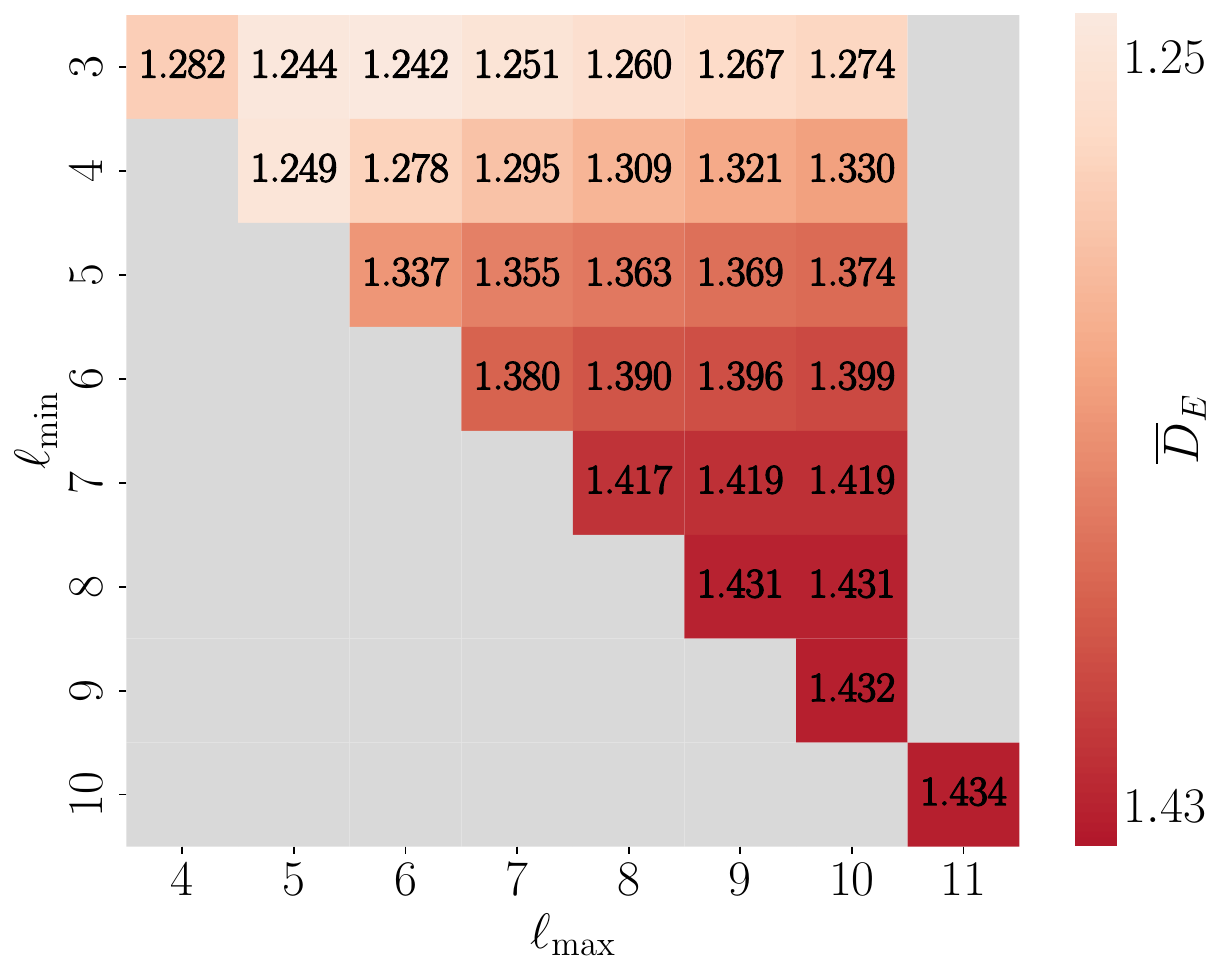}
    \caption{Time-averaged diffusion coefficient $\bar{D}_E$ (averaged over $t\in[20,100]$) as a function of truncation parameters $(\ell_{\min},\ell_{\max})$ for $q_{\max}=0.5{\%}$. The diffusion coefficients vary weakly with $\ell_{\max}$ at fixed $\ell_{\min}$, while increasing $\ell_{\min}$ systematically increases $\bar{D}_E$ and drives it toward a converged value.}
    \label{fig:results_colormap}
\end{figure}

For a subsystem Hilbert space dimension $d_{\ell}=2^{\ell+1}$, following the original LITE conventions, LITE requires a matrix decomposition to evaluate the von~Neumann entropy and therefore scales asymptotically as $\mathcal{O}(d_{\ell}^{3})$ per entropy evaluation, whereas \rlite{} computes the \renyi{} entropy from $\tr(\rho^2)$ using matrix contractions with $\mathcal{O}(d_{\ell}^{2})$ cost. Figure~\ref{fig:benchmarking}(g) confirms this expected polynomial separation in the measured runtime as a function of $d_{\ell_{\max}}$, implying an increasing speed-up with $\ell_{\max}$. To put these dimensions in perspective, the largest value shown, $d_{\ell_{\max}}=2^{\ell_{\max}+1}\simeq 8192$, corresponds to reduced density matrices on $(\ell_{\max}+1)=13$ spin-$1/2$ degrees of freedom. At this size, \rlite{} is already faster by roughly six orders of magnitude compared to LITE.

We next verify that this replacement does not change the transport physics in the standard mixed-field Ising benchmark. Figure~\ref{fig:benchmarking}(a)--(f) compares the diffusion coefficient $D_E(t)$ obtained with LITE and \rlite{} for increasing $\ell_{\min}$ at fixed $\ell_{\max}=9$ and $q_{\max}=0.5\%$. Across the full range $\ell_{\min}=3,\dots,8$, the two methods produce nearly identical time dependence of the diffusion coefficient, with discrepancies that remain small on the scale relevant for extracting the late-time diffusion plateau. 
This shows that replacing the von~Neumann information measure by its \renyi{} counterpart leaves the transport dynamics essentially unchanged over a broad range of truncation parameters.

At the same time, one should not expect the same numerical value of $q_{\max}$ to play exactly the same role in LITE and \rlite{}. The information-removal step is triggered by the amount of accumulated local information at the top level, and because the two algorithms quantify this information differently, a given $q_{\max}$ can correspond to slightly different minimization times and therefore to a slightly different extracted value of $\bar{D}_E$. In \rlite{} this threshold is applied to $\gamma_{n,\mathrm{pos}}^\ell=\max(\gamma_n^\ell,0)$, so negative signed purity-gain contributions are monitored but cannot cancel positive top-level buildup. For this reason, the meaningful comparison is not at a single matched value of $q_{\max}$, but in the regime where $\bar{D}_E$ becomes insensitive to $q_{\max}$. The inset of Fig.~\ref{fig:benchmarking}(f) shows that such a plateau exists. Within this stability window, $\bar{D}_E$ depends only weakly on $q_{\max}$, and LITE and \rlite{} agree within the uncertainty relevant for extracting the diffusion plateau. This agreement shows that, once the threshold is chosen in the converged regime, the extracted diffusion constant is insensitive to the precise implementation of the information measure.

\begin{figure}[t]
    \centering
    \includegraphics[width=\linewidth]{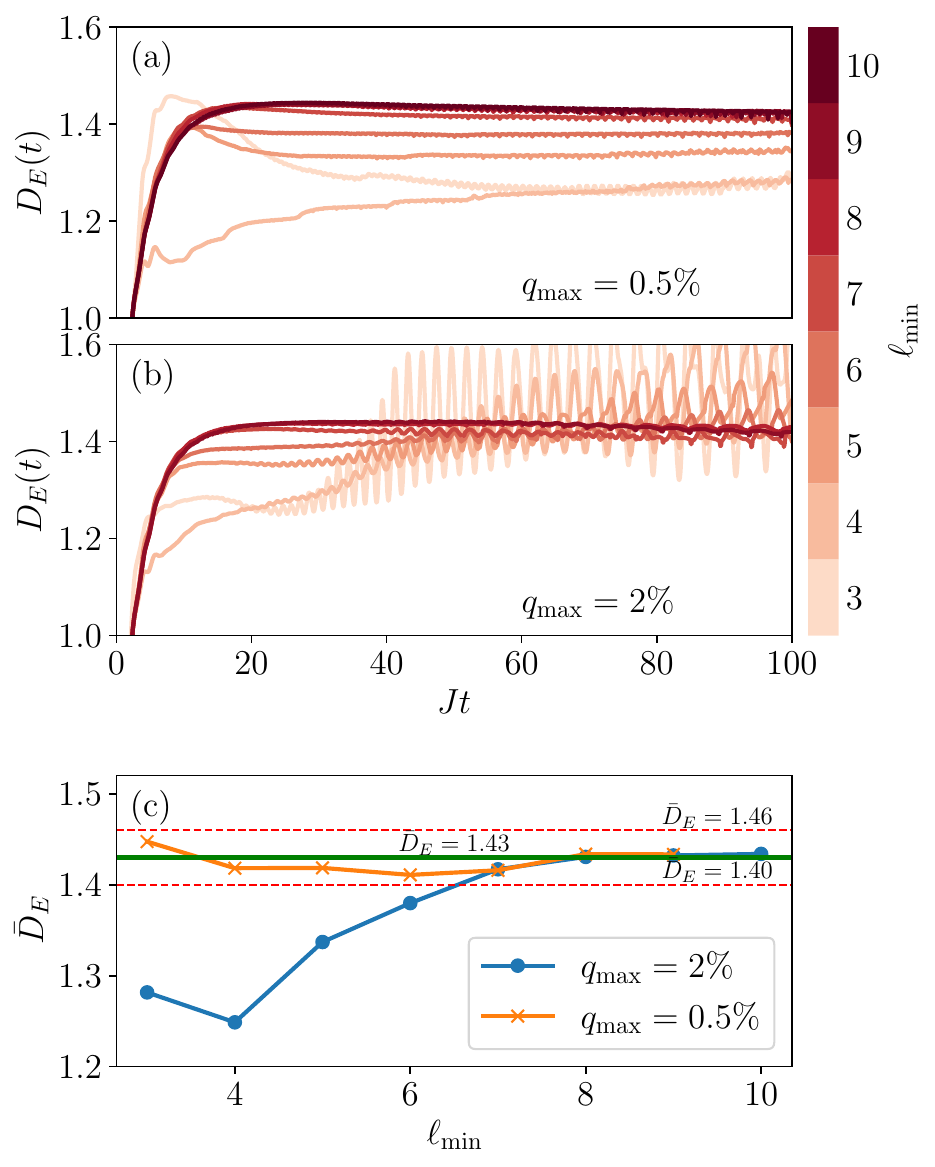}
    \caption{Time-dependent diffusion coefficient $D_E(t)$ for increasing $\ell_{\min}$ at fixed $\ell_{\max}=\ell_{\min}+1$, shown for two information-removal thresholds: (a) $q_{\max}=0.5\%$ and (b) $q_{\max}=2.0\%$. At late times (here $t\gtrsim 20$), a diffusive plateau emerges and becomes progressively more stable as $\ell_{\max}$ is increased, enabling a more reliable estimate of the diffusion constant. Panel (c) highlights the convergence of the extracted plateau value $\bar{D}_E$ with $\ell_{\min}$ and demonstrates its robustness with respect to varying $q_{\max}$.}
    \label{fig:results_dt_plateau}
\end{figure}

\subsubsection{Larger $\ell_{\max}$ and improved extrapolation of late-time transport.} 
The main practical consequence of the reduced run time is that \rlite{} gives access to larger truncation scales $\ell_{\max}$ than standard LITE. This is essential because the two LITE cutoffs play different roles. The upper scale $\ell_{\max}$ determines how far information can propagate up the hierarchy before the closure of \eqref{eq:subsystem_eom} is forced with a recovery step, whereas $\ell_{\min}$ sets the scale at which the information-removal step can feed back on the physical dynamics. Accordingly, once $\ell_{\max}$ is sufficiently large, the extracted diffusion constant depends much more strongly on $\ell_{\min}$ than on $\ell_{\max}$, as shown in Fig.~\ref{fig:results_colormap}. In this sense, convergence of the dynamics produced by LITE should primarily be assessed with respect to $\ell_{\min}$, while using $\ell_{\max}$ to ensure that this extrapolation is not itself cutoff limited (as we discuss in Sec.~\ref{subsec:floquet}).

This implies that the natural convergence strategy is to increase $\ell_{\min}$ while keeping $\ell_{\max}$ as close as possible above it. In our data this is implemented by the near-diagonal choice $\ell_{\max}=\ell_{\min}+1$. Along this line, increasing $\ell_{\min}$ simultaneously pushes the closure scale to larger subsystems and delays the feedback of information removal. Convergence with respect to $\ell_{\min}$ then means that successive near-diagonal points yield the same late-time plateau value $\bar{D}_E$. Panels (a) and (b) of Fig.~\ref{fig:results_dt_plateau} make this behavior explicit. For small $\ell_{\min}$, the late-time plateau is systematically underestimated because the hierarchical closure truncates the flow of purity to larger length scales too early. As a consequence, the long-wavelength hydrodynamic mode is incompletely represented, the energy packet broadens too slowly, and $D_E(t)=\tfrac{1}{2}\partial_t\sigma_E^2(t)$ is reduced. As $\ell_{\min}$ and $\ell_{\max}$ are increased together, this cutoff effect is reduced and the late-time curves collapse onto a common plateau.

With \rlite{} we can push this procedure to $\ell_{\max}=11$, for which the plateau has clearly stabilized and no extrapolation in $\ell_{\min}$ is required. Figure~\ref{fig:results_dt_plateau}(c) further shows that the corresponding values of $\bar{D}_E$ are stable for $q_{\max}=0.5\%$ and $q_{\max} = 2\%$. In contrast, Ref.~\cite{artiacoEfficientLargeScaleManyBody2024} was limited to $\ell_{\max}\le 9$ and therefore estimated the asymptotic diffusion constant by extrapolating $\bar{D}_E$ linearly in $1/\ell_{\min}$ to $\ell_{\min}\to\infty$, obtaining $\bar{D}_E=1.55$ \cite{artiacoEfficientLargeScaleManyBody2024}. Our converged value, $\bar{D}_E\approx 1.43$, lies within the range of values reported in the literature: $\bar{D}_E\approx 1.40$--$1.46$ \cite{Pollmann2022,thomas2023comparing,wang2023diffusion,Parker_2019}. Importantly, the largest explicitly simulated values in Ref.~\cite{artiacoEfficientLargeScaleManyBody2024} are already consistent with this plateau, so the remaining discrepancy can be attributed to the extrapolation rather than to the underlying finite-$\ell$ data. In this sense, the larger $\ell_{\max}$ accessible with \rlite{} substantially reduces the reliance on an extrapolation in $\ell_{\min}$ and yields a diffusion constant that is consistent with the literature range.

\begin{figure*}[t]
    \centering
    \includegraphics[width=\textwidth]{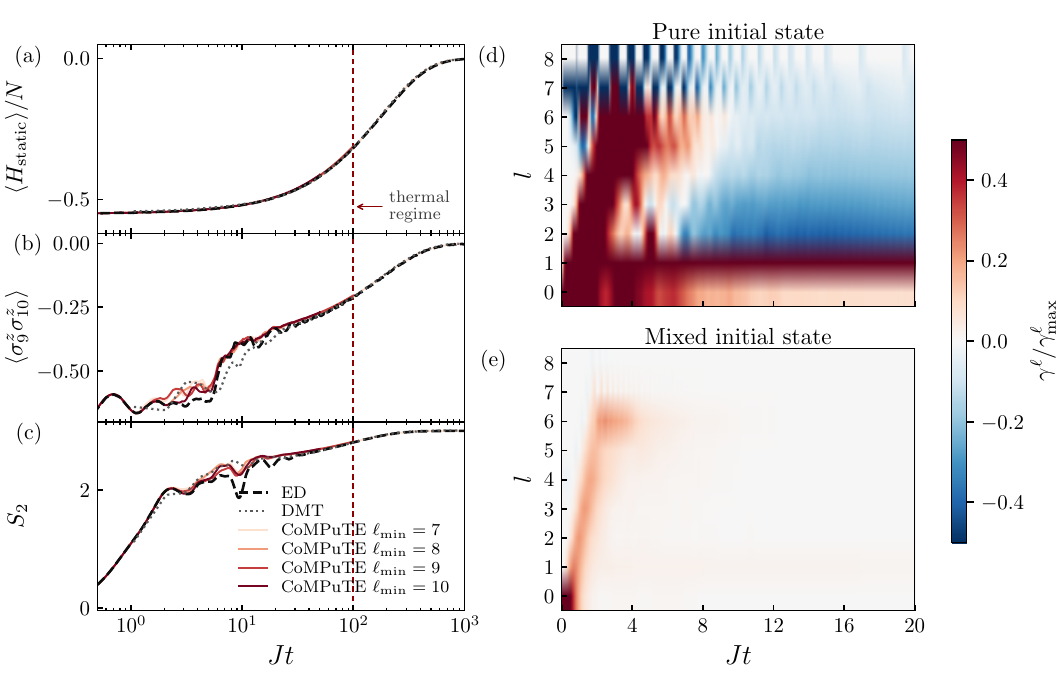}
    \caption{Driven Floquet dynamics for a chain with $N=20$. Panels (a)--(c) compare ED (black dashed), DMT (black dotted) \cite{whiteQuantumDynamicsThermalizing2018}, and \rlite{} for increasing values of $\ell_{\min}$ (solid colors) using the static energy density $\langle H_{\text{static}}\rangle/N$, the local correlator $\langle \sigma^z_9 \sigma^z_{10}\rangle$, and the second R\'enyi entropy $S_2$ of the leftmost three sites. The red dashed line marks the onset of the thermal regime. In panels (a)--(c), the late-time regime was simulated explicitly only with \rlite{} at $\ell_{\min}=7$, because simulations with larger $\ell_{\min}$ already capture the transient regime, establishing that they also reproduce the correct late-time behavior. Panels (d) and (e) show the level-resolved purity gain normalized by the maximum absolute value $\gamma_{\mathrm{max}}$, for the pure state in \eqref{eq:floquet_dw_neel_pure} and the locally mixed state in \eqref{eq:floquet_dw_neel_mixed} with $m=0.2$, respectively. The pure initial state generates strong positive and negative scale-resolved purity gain that spreads to higher levels and then decays during heating, whereas the mixed initial state produces only a weak, short-lived response.}
    \label{fig:floquet_info_flow}
\end{figure*}

\subsection{Driven Floquet dynamics}
\label{subsec:floquet}

We next test \rlite{} on a driven (Floquet) spin chain and compare to DMT \cite{whiteQuantumDynamicsThermalizing2018}, which provides a widely used baseline for long-time non-equilibrium dynamics in one dimension. We primarily use a pure product initial state, which would be prohibitive in the standard LITE formulation but is amenable for \rlite{}, and compare its purity-gain flow to a locally mixed variant defined below.

In the generic case, a non-integrable Floquet system is expected to \emph{heat} under periodic driving, relaxing toward an infinite-temperature state at late times (up to possible long-lived prethermal regimes at high drive frequencies). This heating is directly reflected in energy-density observables and in the growth and saturation of subsystem entropies.

We consider a chain of length $L=N=20$. With the convention $1\equiv\uparrow$ and $0\equiv\downarrow$, the domain-wall N\'eel pattern with a domain wall every four sites is
\begin{equation}
    b_i=\frac{1+(-1)^{i+\lfloor i/4\rfloor}}{2},
    \label{eq:floquet_dw_neel_bits}
\end{equation}
The pure initial state is therefore
\begin{equation}
    \ket{\psi_{\text{init}}}
    =\bigotimes_{i=0}^{L-1}\ket{b_i}
    =\ket{10100101\cdots}.
    \label{eq:floquet_dw_neel_pure}
\end{equation}
For comparison in the purity-gain analysis, we also use the locally mixed product state
\begin{equation}
\rho_{\text{init}}(m)
=
\bigotimes_{i=0}^{L-1}
\frac{1}{2}
\left[
\id
+
m\,(-1)^{i+\lfloor i/4\rfloor}\sigma_i^z
\right],
\qquad 0\le m\le 1,
\label{eq:floquet_dw_neel_mixed}
\end{equation}
with $m=0.2$. We evolve under a time-periodic Hamiltonian $H(t)=H_{\text{static}}+H_{\text{drive}}(t)$ with
\begin{equation}
\begin{aligned}
H_{\text{static}}
&=
\sum_{m=1}^{N-1}\Big[
J \sigma^z_m \sigma^z_{m+1}
+ J_x \sigma^x_m \sigma^x_{m+1}
\Big]
+ h_x \sum_{m=1}^{N} \sigma^x_m,
\\
H_{\text{drive}}(t)
&=
\sum_{m=1}^{N}
\mathrm{sgn}\!\big[\cos(\omega t)\big]\,
\Big(h_y \sigma^y_m + h_z \sigma^z_m\Big),
\end{aligned}
\end{equation}
with $\omega=6J$ and $(J,J_x,h_x,h_y,h_z)=(1,0.75,0.21,0.17,0.13)$, so that the drive period is $T=2\pi/\omega$.

Fig.~\ref{fig:floquet_info_flow} compares DMT data from Ref.~\cite{whiteQuantumDynamicsThermalizing2018}, and \rlite{} to reference ED calcution of the static energy density $\langle H_{\text{static}}\rangle/N$, a representative local correlator $\langle \sigma^z_9 \sigma^z_{10}\rangle$, and the \renyi{} entropy $S_2$ of the leftmost three sites. 
The observed drift of $\langle H_{\text{static}}\rangle/N$ toward its infinite-temperature value is perfectly captured by all methods, and for \rlite{} we have computed the late time asymptote only for $\ell_{\min=7}$, because it captures the behavior already perfectly.
The local spin correlator and $S_2$ on the three leftmost edge sites show a more nuanced behavior. The approximate methods again reproduce the correct asymptotic trend, but the early-time regime is more challenging. For these observables, DMT mainly captures the overall trend, whereas \rlite{} resolves the early-time dynamics systematically better as $\ell_{\min}$ is increased. The remaining discrepancies are mostly found in the fine structure of the intermediate-time regime.

In Fig.~\ref{fig:floquet_info_flow}(d) and (e), we demonstrate the diagnostic capabilities of the scale-resolved purity gain $\gamma^\ell$ for the pure and locally mixed initial states, respectively.
These diagnostics reveal a clear distinction between the two initial conditions.
Starting from the pure product state, Fig.~\ref{fig:floquet_info_flow}(d) shows a strong signed purity-gain transient that rapidly spreads across the retained hierarchy.
Because the purity gain is signed, both positive and negative values can occur for physical states.
In particular, the alternating positive and negative pattern at the top retained level is a direct consequence of the \rlite{} closure, where triggered information removal performs a purity minimization on the largest retained subsystems.
The pronounced imbalance between negative and positive top-level purity gain is induced by the minimum-purity bias of this information-minimization step.
In contrast, the locally mixed initial state in Fig.~\ref{fig:floquet_info_flow}(e), evolved under the same drive, produces only a weak and short-lived response on the same scale. The extreme top-level alternation is absent because, for this initial condition, the minimum-purity update is closer to the true optimum of the loss function \eqref{eq:chi_leastsquares}.
Panel (d) shows that the largest redistribution of purity gain occurs for $Jt\lesssim 8$, the same early-time regime in which \rlite{} captures the local observables. The remaining discrepancies at intermediate times are comparatively featureless in the purity-gain evolution, which suggests that they are artifacts of the finite subsystem-hierarchy truncation rather than an uncontrolled instability of the method.

\subsection{Superdiffusion in the XXZ spin chain}
\begin{figure}[t]
    \centering
    \includegraphics[width=\linewidth]{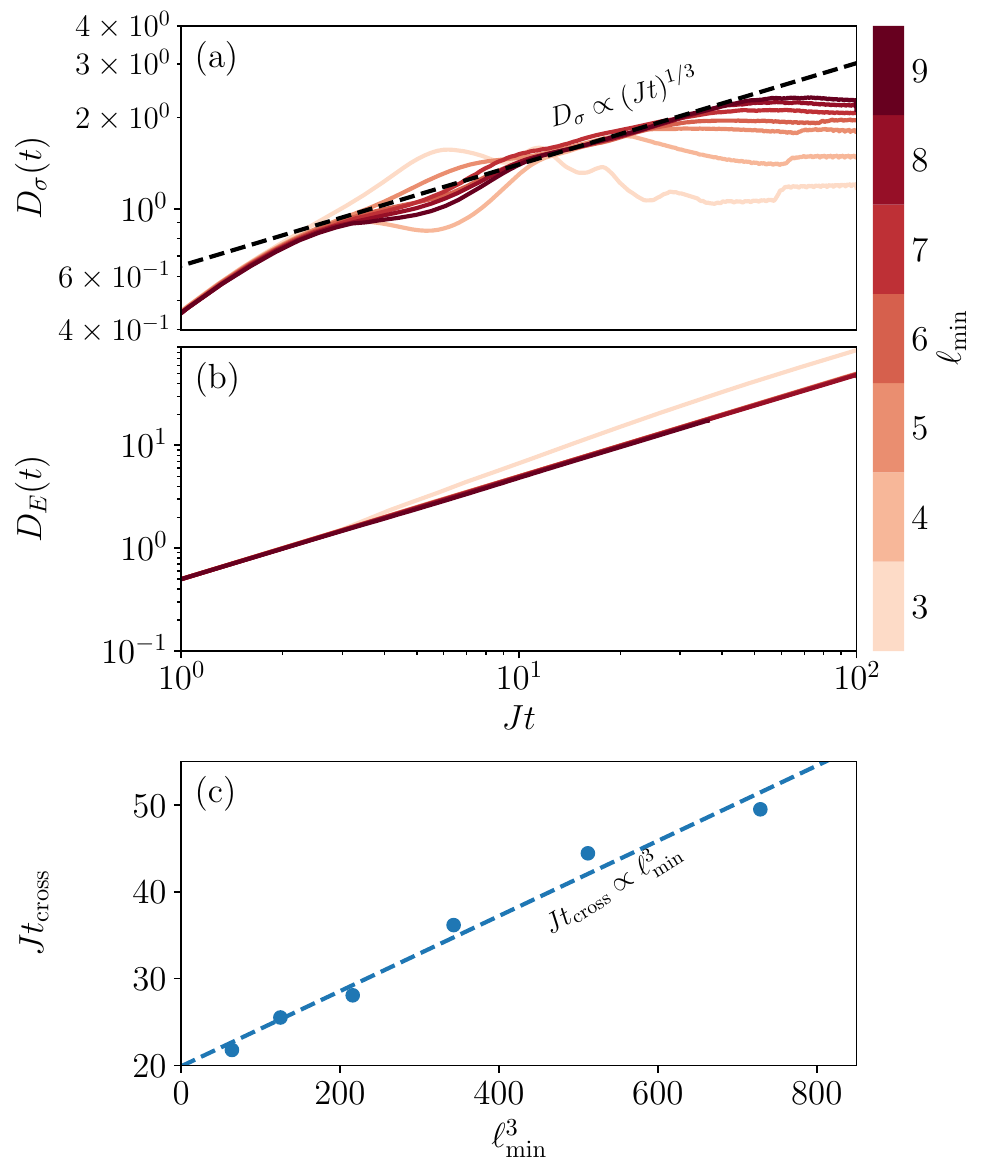}
    \caption{
    (a) Time dependent spin diffusion coefficient $D_\sigma(t)$ at $\Delta=1$ obtained with \rlite{} for $\ell_{\max}=\ell_{\min}+1$ and truncation threshold $q=0.5\%$. The dashed line indicates the expected superdiffusive scaling $D_\sigma(t)\propto (Jt)^{1/3}$ \cite{gopalakrishnanKineticTheorySpin2019a}. As $\ell_{\min}$ is increased the superdiffusive window extends to later times, while a breakdown occurs at a crossover time $t_{\rm cross}$ set by the finite operator support.
    (b) Time dependent energy transport coefficient $D_E(t)$ at $\Delta=1$ for the same \rlite{} parameters. The sustained growth is consistent with ballistic energy transport and a nonzero Drude contribution in the energy channel.
    (c) Crossover times $t_{\rm cross}$ extracted from panel (a) plotted against $\ell_{\min}^3$. The approximately linear dependence supports the kinetic theory estimate $Jt_{\rm cross}\propto \ell_{\min}^3$ that follows from a string cutoff $s^\ast(t)\propto (Jt)^{1/3}$ together with $s_{\max}\propto \ell_{\min}$ \cite{gopalakrishnanKineticTheorySpin2019a}.}
    \label{fig:neel}
\end{figure}
Integrable quantum systems host an extensive number of conserved quantities which constrain current relaxation and can give rise to hydrodynamic signatures on all length scales. These properties constitute an interesting and relevant stress test for \rlite{}. We consider the spin-1/2 XXZ chain
\begin{equation}
    H = J \sum_{m=1}^{N-1} \Bigl[ S^x_m S^x_{m+1} + S^y_m S^y_{m+1} + \Delta S^z_m S^z_{m+1} \Bigr],
    \label{eq:xxz}
\end{equation}
where $S^\mu_m$ are local spin operators and $J=1$. For $\Delta/J<1$ spin transport has a nonzero Drude weight and supports ballistic spin transport, while for $\Delta/J>1$ the spin Drude weight vanishes and spin transport is diffusive at high temperature. Precisely at $\Delta/J=1$ the spin Drude weight still vanishes, yet transport is anomalous and becomes superdiffusive \cite{gopalakrishnanKineticTheorySpin2019a}. We study $\Delta/J=1$ with \rlite{} to highlight its ability to stably propagate larger subsystem density matrices in a regime where superdiffusion is controlled by quasiparticles with growing spatial support.

At $\Delta/J=1$ the expected scaling of the spin diffusion coefficient is
\begin{equation}
    D_\sigma(t) \propto (Jt)^{1/3},
    \label{eq:supdiff_scaling}
\end{equation}
while the energy channel is ballistic and therefore satisfies
\begin{equation}
    D_E(t) \propto Jt,
    \label{eq:ballistic_scaling}
\end{equation}
which is consistent with a nonzero energy Drude weight in the XXZ chain \cite{gopalakrishnanKineticTheorySpin2019a}.

Our initial states are weak local perturbations of the infinite-temperature state. For energy transport we use the same form as in \eqref{eq:initial_state} with $\beta=0.05$, start from $N=9$ with $\rho^{\ell}_{n,\mathrm{init}}$ supported on three central sites, and dynamically enlarge the system so that information does not reach the boundaries; the final size is $N=273$. For spin transport we use $\rho^{\ell}_{n,\mathrm{init}} = \tr[e^{-\epsilon S^z_n}]^{-1} e^{-\epsilon S^z_n}$ with $\epsilon=0.05$, start from $N=9$ with a single perturbed central site, and reach final sizes from $N=120$ to $N=129$ depending on $\ell_{\max}$. We fix the truncation threshold to $q=0.5\%$ and parametrically lock the relevant subsystem density matrix scales $\ell_{\max}=\ell_{\min}+1$.

Fig.~\ref{fig:neel}(a) shows that \rlite{} reproduces the analytic superdiffusive trend $D_\sigma(t)\propto (Jt)^{1/3}$ over a growing time window as $\ell_{\min}$ is increased. The energy counterpart $D_E(t)$ in Fig.~\ref{fig:neel}(b) exhibits the sustained growth consistent with \eqref{eq:ballistic_scaling}. The fact that the sustained growth breaks down in the spin channel but not in the energy channel reflects the different carriers of transport at $\Delta/J=1$. Energy transport is dominated by a ballistic contribution that is tied to comparatively local conserved currents and is therefore robust under \rlite{} truncation. Spin superdiffusion instead is carried by quasiparticles with increasingly large support, which requires growing effective operator support, leading to a truncation-dependent saturation of the diffusion coefficient at late times.

The kinetic theory of Ref.~\cite{gopalakrishnanKineticTheorySpin2019a} attributes \eqref{eq:supdiff_scaling} to \emph{Bethe ansatz string} quasiparticles of length $s$ whose velocities scale as $v_s \propto 1/s$. The effective number of strings that contribute up to time $t$ grows as
\begin{equation}
    s^\ast(t) \propto (Jt)^{1/3},
    \qquad
    D_\sigma(t) \propto s^\ast(t),
    \label{eq:string_cutoff}
\end{equation}
which is another way to state \eqref{eq:supdiff_scaling}. Any finite truncation of the operator manifold imposes a maximal effective string length $s_{\max}$, which in LITE is controlled by $\ell_{\max}$. The superdiffusive regime therefore ends when $s^\ast(t)$ reaches this cutoff, giving the scaling estimate
\begin{equation}
    s^\ast(t_{\rm cross}) \sim s_{\max} \propto \ell_{\min}
    \;\Rightarrow\;
    Jt_{\rm cross} \propto \ell_{\min}^3.
    \label{eq:tcross_scaling}
\end{equation}
We test \eqref{eq:tcross_scaling} by extracting the deviation time $t_{\rm cross}$ at which $D_\sigma(t)$ departs from the $(Jt)^{1/3}$ growth in Fig.~\ref{fig:neel}(a). Fig.~\ref{fig:neel}(c) shows that $Jt_{\rm cross}$ scales linearly with $\ell_{\min}^3$, which supports the interpretation that LITE truncation limits the accessible quasiparticle string length and therefore bounds the duration over which superdiffusion can be observed.

These results also highlight a practical advantage of \rlite{} over LITE for long-time integrable dynamics. Superdiffusion is a cumulative long-time effect that relies on increasingly nonlocal operators, and reaching the corresponding regime requires stable propagation at larger $\ell_{\max}$. \rlite{} improves this effective finite-size extrapolation and makes the intermediate-time superdiffusive window visible, as demonstrated in Fig.~\ref{fig:neel}(a).
However, since LITE propagates the density matrix rather than a single pure state, \emph{exact diagonalization} can access a broader superdiffusive time window by evolving the pure state directly. Further improvements to \rlite{} may overcome this hurdle in the future.

\section{Discussion}
In this work we introduced compressed minimum-purity time evolution---\rlite{}---as an approach to enable numerical simulation of quantum dynamics up to late times by systematically discarding irrelevant information.
Inspired by the LITE algorithm, the method propagates a consistent set of local reduced density matrices in time.
The key conceptual difference is, however, that truncation and recovery to restrict the hierarchical equations of motion are based on a minimum purity principle and employ the conservation of purity currents as a constraint.
Thereby, information removal and the recovery map can be implemented without matrix logarithms or dense matrix decompositions.
As a consequence, the cost of the entropy evaluation is reduced from the cubic scaling of the von~Neumann entropy-based LITE algorithm to a quadratic scaling in the subsystem Hilbert-space dimension.
In practice this yields a speedup that grows with $d_{\ell_{\max}}$, and for the largest subsystem sizes reached in our benchmarks \rlite{} is already faster by roughly six orders of magnitude.
Equally important, the absence of matrix logarithms removes the numerical instability associated with small or vanishing eigenvalues, which makes pure-state propagation practical and allows stable simulations at larger $\ell_{\max}$ than were previously accessible with von~Neumann entropy-based LITE.

We verified that these simplifications do not compromise the transport physics in the regime where LITE is known to perform well.
For the mixed-field Ising model, \rlite{} reproduces the same time-dependent diffusion coefficient as the original LITE implementation over the full accessible time window when the truncation threshold is chosen in the physically balanced regime.
Moreover, the larger subsystem sizes accessible with \rlite{} reduce the sensitivity to the extrapolation in $\ell_{\min}$ and yield a late-time diffusion plateau at the level of precision established by other state-of-the-art benchmarks \cite{Pollmann2022,thomas2023comparing,wang2023diffusion,Parker_2019}.
In this sense, \rlite{} retains the central strength of LITE, namely accurate late-time transport, while substantially extending the numerically accessible truncation scales. 

A second main outcome is that \rlite{} extends the scope of LITE simulations beyond near-thermal mixed-state settings.
In the driven Floquet example, starting from an uncorrelated pure initial state, \rlite{} reproduces the relaxation and heating trends obtained with density matrix truncation across the full investigated time window.
This shows that the method can treat genuinely out-of-equilibrium dynamics and it also demonstrates the ability to propagate pure initial states, which enlarges the class of dynamical problems that can be addressed by LITE-type methods.

The integrable XXZ chain at $\Delta=1$ provides a complementary lesson.
Here \rlite{} puts the core approximation underlying the reduced-density description, in a controlled manner, to the test.
Spin superdiffusion at the isotropic point is a cumulative long-time effect that relies on contributions from increasingly nonlocal operators, and this ultimately exposes the finite-$\ell_{\max}$ limitation of the method.
Accordingly, the superdiffusive window extends systematically as $\ell_{\max}$ is increased, but eventually breaks down at a crossover time set by the maximal operator support that can still be represented.
At the same time, the improved scaling of \rlite{} makes this intermediate-time regime visible and systematically improvable through $\ell_{\max}$ extrapolation.
Energy transport in the same model displays its expected sustained growth over the accessible window, which further highlights that the breakdown is tied to the structure of the relevant transport modes rather than to a generic instability of the algorithm.

These observations help clarify both why the LITE approach works and where its limits lie.
When late-time behavior is governed by information that remains encoded in reduced density matrices of moderate size, controlled information removal can bypass the entanglement barrier and still preserve the observables of interest.
When the relevant dynamics depends on operator structures whose support keeps growing, the truncation in subsystem size becomes the dominant limitation.
Even in that case, however, \rlite{} turns the limitation into a quantitative convergence problem in $\ell_{\max}$ rather than an immediate numerical breakdown, which is precisely what makes the XXZ example so instructive.

Finally, the scale-resolved purity gain constitutes a new diagnostic tool to characterize the dynamics of quantum many-body systems in terms of information flow.
While we find that its dynamics closely resembles the local information on the information lattice, the cost of computing the purity gain is reduced in comparison to von~Neumann entropy---analogously to the truncation and recovery steps.
Analyzing the behavior of purity gain under quantum dynamics in more detail is a task for future research.

Looking further ahead, the enlarged subsystem sizes accessible via \rlite{} suggest that initial applications to higher-dimensional settings may become less prohibitive.
Extending the information-lattice strategy beyond one dimension remains speculative and will require new ideas in geometry \cite{florHigherDimensionalInformationLattice2025}, recovery, and truncation.
While the present reduction in computational cost removes an important practical obstacle and provides a more favorable starting point for such developments, further scaling of the method may also require addressing the memory cost.
More broadly, our results show that the information-lattice approach is flexible enough to accommodate different entropy measures and closure strategies for efficient time evolution.
This combination of improved efficiency, access to pure-state dynamics, and scale-resolved purity gain diagnostics makes \rlite{} a promising basis for future studies of late-time quantum dynamics and transport.

\section*{Author contributions} 
M.B. developed and implemented the CoMPuTE algorithm, including the algorithmic improvements that constitute the central methodological contribution of this work. 
He performed the numerical simulations, produced and analyzed the data and figures. J.B.R. formalized the mathematical derivations underlying the methodological developments, directly supervised the algorithmic and numerical work, analyzed and interpreted the data, and wrote the manuscript. M.S. conceived and supervised the overall project. All authors continuously discussed the results and contributed to the development of the manuscript. 

A large language model was used to assist with grammar and language editing and to check the presentation and internal consistency of mathematical derivations and formulas in the final manuscript. All resulting suggestions were independently assessed by the authors, who take full responsibility for the content and correctness of the work.

\begin{acknowledgments}
The authors thank Claudia Artiaco for insightful feedback and inspiring discussions that helped to develop the presentation of the results.
The authors acknowledge insightful discussions with Jens H. Bardarson, Sourav Nandy and Marin Bukov.
This work was supported via the Helmholtz Initiative and Networking Fund, grant no.~VH-NG-1711.
The authors gratefully acknowledge the Gauss Centre for Supercomputing e.V. (www.gauss-centre.eu) for funding this project by providing computing time through the John von Neumann Institute for Computing (NIC) on the GCS Supercomputer JUWELS at Jülich Supercomputing Centre (JSC).
\end{acknowledgments}

\begin{appendix}
\onecolumngrid
\section{\renyi{} enabled local information removal}
\label{app:proofs1}

During the hierarchical time evolution of the reduced density matrices information is free to spread to greater length scales. The maximum system size curbs this spread, leading to finite-size effects and eventual failure to produce local Gibbs states. In Ref.~\cite{artiacoEfficientLargeScaleManyBody2024}, Artiaco et al.\ introduced a well-controlled protocol to counteract the buildup of information. Central to this protocol is the LITE ``information minimization'' step that removes excess information from the working-level state
$\rho_n^{\ell}$ by adding an allowed correction $\chi_n^{\ell}$,
\begin{equation}
	\rho_n^{\ell}\;\mapsto\;\rho_n^{\ell}+\chi_n^{\ell}.
\end{equation}
The correction is required to preserve the left and right boundary reductions and the information currents across the left/right boundary links on the information lattice. For \rlite{} these currents are the R\'enyi-2 purity currents defined in Eqs.~\xeqref{eq:renyi_currents_intro_L} and \xeqref{eq:renyi_currents_intro_R}; they remain meaningful boundary-current quantities even though the induced scale-resolved purity gain is signed. These conditions can be formulated as linear constraints. First, the left and right marginals of $\rho^\ell_n$ are preserved if the left and right partial traces of $\chi_n^{\ell}$ are zero,
\begin{equation}
    \tr_L(\chi_n^{\ell}) = \tr_R(\chi_n^{\ell}) = 0\;.
\end{equation}
Similarly, the information currents to the left and right links across the level boundary $\ell \leftrightarrow \ell - 1$ need to be unaffected by $\chi_n^{\ell}$, which translates to
\begin{equation}
    J_L(\chi_n^{\ell}) = J_R(\chi_n^{\ell}) = 0\;.
\end{equation}
The information currents are defined as
\begin{align}
    J_L(\rho_n^{\ell}) = -i\tr\Big(\big[ \nabla_\rho S^\ell_n, H^{\ell+1}_{n - 1/2} - H^\ell_n \big]\rho^{\ell+1}_{n-1/2}\Big)\;,\\
    J_R(\rho_n^{\ell}) = -i\tr\Big(\big[ \nabla_\rho S^\ell_n, H^{\ell+1}_{n + 1/2} - H^\ell_n \big]\rho^{\ell+1}_{n+1/2}\Big) \; ,
\end{align}
where the entropy gradient is derived from the \renyi{} entropy
\begin{equation}
    \nabla_\rho S^\ell_n = - \frac{2}{\tr\big[(\rho^\ell_n)^2\big]} \big(\rho^\ell_n\big)^T\;.
\end{equation}

All of these are \emph{linear} constraints on $\chi^\ell_n$.
Abstractly, we collect them into a single linear map
\begin{equation}
	\mathcal{B}_{\mathrm{IR}}(\chi)\equiv
	\bigl(\tr_L(\chi),\;\tr_R(\chi),\;J_L(\chi),\;J_R(\chi)\bigr) \; ,
\end{equation}
and impose
\begin{equation}
	\mathcal{B}_{\mathrm{IR}}(\chi)=0 \;.
	\label{eq:BIR_constraints}
\end{equation}
The feasible corrections therefore form a linear subspace
\begin{equation}
	\mathcal{V}\equiv\ker\mathcal{B}_{\mathrm{IR}}.
\end{equation}

In the original formulation $\chi^\ell_n$ is determined by a constrained maximum-entropy problem,
\begin{equation}
	\chi_n^{\ell}=\arg\max_{\chi\in\mathcal{V}} S(\rho_n^{\ell}+\chi),
	\label{eq:chi_maxentropy_generic}
\end{equation}
with $S$ an entropy measure. With our R\'enyi-2 choice, the same step becomes a simple minimal purity problem
\begin{equation}
	\chi_n^{\ell}=\arg\max_{\chi\in\mathcal{V}} S_2(\rho_n^{\ell}+\chi)
	\;=\;\arg\min_{\chi\in\mathcal{V}} \tr\!\bigl[(\rho_n^{\ell}+\chi)^2\bigr].
	\label{eq:chi_maxS2_minpurity}
\end{equation}
Using an alternative formulation of the minimal purity problem
\begin{equation}
    \label{eq:minimum_purity_problem}
    \arg\min_\rho \tr(\rho^2) = \arg\min_\rho \left\| \rho - \frac{\id}{d}\right\|^2_2
\end{equation}
and noting that $\tr(\rho_n^{\ell}+\chi)=1$
for all $\chi\in\mathcal{V}$, the optimization \eqref{eq:chi_maxS2_minpurity} is equivalent to
\begin{equation}
	\chi_n^{\ell}=\arg\min_{\chi\in\mathcal{V}}
	\left\|\rho_n^{\ell}+\chi-\frac{\mathbb{I}}{d}\right\|_2^2.
	\label{eq:chi_leastsquares}
\end{equation}
But \eqref{eq:chi_leastsquares} is exactly a least-squares problem. To zeroth order we obtain the solution of \eqref{eq:chi_leastsquares} by orthogonally projecting the vector
$\frac{\mathbb{I}}{d}-\rho_n^{\ell}$ onto $\mathcal{V}$
\begin{equation}
		\;\tilde\chi_n^{\ell}=\frac{\mathbb{I}}{d}-\rho_n^{\ell},
		\qquad
		\chi_n^{\ell}=\mathcal{P}_{\mathcal{V}}\!\bigl(\tilde\chi_n^{\ell}\bigr)\;.
	\label{eq:chi_zerothorder}
\end{equation}
The explicit trace- and current-preserving projector used for the information-removal constraint $\mathcal{B}_{\mathrm{IR}}(\chi)=0$ follows Eqs.~(21)--(29) of Ref.~\cite{artiacoEfficientLargeScaleManyBody2024}.
Given an arbitrary matrix $\chi_n^{\ell}$, the projectors onto the left and right partial-trace-free spaces are
\begin{align}
	P^{\mathrm{Tr}}_{L}\chi_n^{\ell}
	&:=
	\chi_n^{\ell}
	-
	\frac{\id_d}{d}\otimes\tr^1_L(\chi_n^{\ell}),
	\label{eq:ir_trace_projector_left}\\
	P^{\mathrm{Tr}}_{R}\chi_n^{\ell}
	&:=
	\chi_n^{\ell}
	-
	\tr^1_R(\chi_n^{\ell})\otimes\frac{\id_d}{d}.
	\label{eq:ir_trace_projector_right}
\end{align}
The left current of the trace-free matrix is
\begin{equation}
	J^r_L(P^{\mathrm{Tr}}_{R}P^{\mathrm{Tr}}_{L}\chi_n^{\ell})
	=
	-i\tr\!\left(
	\left[
	\nabla_{\rho}S^{\ell-r}_{n+r/2},
	H_n^{\ell}-H^{\ell-r}_{n+r/2}
	\right]
	P^{\mathrm{Tr}}_{R}P^{\mathrm{Tr}}_{L}\chi_n^{\ell}
	\right).
	\label{eq:ir_left_current_projected}
\end{equation}
Using Hermiticity and cyclicity of the trace, this can be written as
\begin{equation}
	J^r_L(P^{\mathrm{Tr}}_{R}P^{\mathrm{Tr}}_{L}\chi_n^{\ell})
	=
	-i\tr(f_n^{\ell}\chi_n^{\ell}),
	\label{eq:ir_left_current_linear}
\end{equation}
where
\begin{equation}
	f_n^{\ell}
	:=
	P^{\mathrm{Tr}}_{R}P^{\mathrm{Tr}}_{L}
	\left[
	\nabla_{\rho}S^{\ell-r}_{n+r/2},
	H_n^{\ell}-H^{\ell-r}_{n+r/2}
	\right].
	\label{eq:ir_left_current_generator}
\end{equation}
Combining the trace-free projections with the projector onto the kernel of $J^r_L$ gives
\begin{equation}
	\chi_{0,n}^{\ell}
	:=
	P^{\mathrm{Tr}}_{R}P^{\mathrm{Tr}}_{L}\chi_n^{\ell}
	\label{eq:ir_trace_projected_chi}
\end{equation}
and
\begin{equation}
	P^J_LP^{\mathrm{Tr}}_{R}P^{\mathrm{Tr}}_{L}\chi_n^{\ell}
	=
	\chi_{0,n}^{\ell}
	-
	\frac{\tr(f_n^{\ell}\chi_{0,n}^{\ell})}{\tr[(f_n^{\ell})^2]}f_n^{\ell}
	\equiv \chi_{1,n}^{\ell}.
	\label{eq:ir_left_current_projector}
\end{equation}
The remaining right current is
\begin{equation}
	J^r_R(\chi_{1,n}^{\ell})
	=
	-i\tr(g_n^{\ell}\chi_{1,n}^{\ell}),
	\label{eq:ir_right_current_linear}
\end{equation}
with
\begin{equation}
	g_n^{\ell}
	:=
	P^J_LP^{\mathrm{Tr}}_{R}P^{\mathrm{Tr}}_{L}
	\left[
	\nabla_{\rho}S^{\ell-r}_{n-r/2},
	H_n^{\ell}-H^{\ell-r}_{n-r/2}
	\right].
	\label{eq:ir_right_current_generator}
\end{equation}
Thus, with the preceding projections understood successively, the concatenated projector onto the kernel of $\mathcal{B}_{\mathrm{IR}}$ is
\begin{equation}
	P_{\mathcal{V}}\big(\chi_n^{\ell}\big)
	\equiv
	\chi_{1,n}^{\ell}
	-
	\frac{\tr(g_n^{\ell}\chi_{1,n}^{\ell})}{\tr[(g_n^{\ell})^2]}g_n^{\ell}.
	\label{eq:ir_total_projector}
\end{equation}

Finally, it is important to note that the approximate solution presented in \eqref{eq:chi_zerothorder} for \eqref{eq:chi_leastsquares} enforces the linear marginal and current constraints, but does not impose positive semidefiniteness of $\rho_n^{\ell}+\chi_n^{\ell}$.

\section{\renyi{} enabled Petz recovery map}
\label{app:proofs2}

In the recovery routine we seek a tripartite state $\rho_{ABC}$ consistent with two overlapping
marginals $\rho_{AB}$ and $\rho_{BC}$ given from lower levels of the information lattice  across the level boundary $\ell \leftrightarrow \ell + 1$,
\begin{equation}
	\tr_C(\rho^{\ell+1}_{ABC})=\rho^\ell_{AB},
	\qquad
	\tr_A(\rho^{\ell+1}_{ABC})=\rho^\ell_{BC},
	\label{eq:marginal_constraints_ABC} 
\end{equation}
for simplicity we drop the $\ell$ index going forward in this section. As in the original LITE construction, we choose the \emph{maximum-entropy} consistent state.
With R\'enyi-2 this reads
\begin{equation}
	\rho_{ABC}
	=\arg\max_{\sigma_{ABC}} S_2(\sigma_{ABC})
	\quad \text{such that}\quad
	\eqref{eq:marginal_constraints_ABC}.
	\label{eq:abc_maxS2}
\end{equation}
By \eqref{eq:minimum_purity_problem} this is equivalently
\begin{equation}
	\rho_{ABC}
	=\arg\min_{\sigma_{ABC}}
	\left\|\sigma_{ABC}-\frac{\mathbb{I}}{d_{ABC}}\right\|_2^2
	\quad \text{such that}\quad
	\eqref{eq:marginal_constraints_ABC}.
	\label{eq:abc_projection_form}
\end{equation}
In the following we prove by construction that the solution to \eqref{eq:abc_projection_form} is the projection of the maximally mixed state onto the affine constraint set \eqref{eq:marginal_constraints_ABC}.

For the construction of the solution to \eqref{eq:abc_projection_form} we need an inner product of matrices, for which we use the Hilbert--Schmidt (HS) inner product
\begin{equation}
	\langle X,Y\rangle \equiv \tr(X^\dagger Y),
	\qquad
	\|X\|_2^2 \equiv \langle X,X\rangle=\tr(X^\dagger X),
\end{equation}
and we restrict to Hermitian matrices (so $X^\dagger=X$).
Next we define the affine constrained set using the linear map 
\begin{equation}
    \mathcal{B}(\sigma)\equiv \big(\tr_C\sigma, \tr_A \sigma \big)
\end{equation}
and the affine offset $b\equiv (\rho_{AB}, \rho_{BC})$, such that the affine constrained set is $\lbrace \sigma: \mathcal{B}(\sigma) = b \rbrace$. With these definitions the problem \eqref{eq:abc_projection_form} can be cast into a Lagrangian with Lagrange multipliers $(\Lambda_{AB},\Lambda_{BC})$
\begin{equation}
    \mathcal{L}(\sigma,\Lambda)
    =
    \frac{1}{2}\left\|\sigma-\frac{\mathbb{I}}{d_{ABC}}\right\|_2^2
    +
    \langle \Lambda,\mathcal{B}(\sigma)-b\rangle_{\oplus},
    \qquad
    \Lambda \,,
    \label{eq:recovery_lagrangian}
\end{equation}
where the codomain $\mathrm{Herm}(AB)\oplus \mathrm{Herm}(BC)$ ($\mathrm{Herm}(X)$ being the vector space of Hermitian operators acting on the system $X$) is equipped with the product HS inner product
\begin{equation}
    \langle (X_{AB},X_{BC}),(Y_{AB},Y_{BC})\rangle_{\oplus}
    \equiv
    \tr(X_{AB}Y_{AB})+\tr(X_{BC}Y_{BC}).
\end{equation}
Stationarity with respect to $\sigma$ gives
\begin{equation}
    0=\nabla_{\sigma}\mathcal{L}
    =
    \sigma-\frac{\mathbb{I}}{d_{ABC}}+\mathcal{B}^{\dagger}\Lambda,
\end{equation}
hence
\begin{equation}
    \sigma
    =
    \frac{\mathbb{I}}{d_{ABC}}-\mathcal{B}^{\dagger}\Lambda.
    \label{eq:sigma_stationary}
\end{equation}
Imposing the constraint $\mathcal{B}(\sigma)=b$ then yields
\begin{equation}
    \mathcal{B}\mathcal{B}^{\dagger}\Lambda
    =
    \mathcal{B}\!\left(\frac{\mathbb{I}}{d_{ABC}}\right)-b.
\end{equation}
Since the two marginal constraints overlap on subsystem $B$, the operator $\mathcal{B}\mathcal{B}^{\dagger}$ is in general only invertible on the constraint range. It is therefore natural to write the solution using the Moore--Penrose pseudoinverse,
\begin{equation}
    \Lambda
    =
    (\mathcal{B}\mathcal{B}^{\dagger})^{+}
    \left[
        \mathcal{B}\!\left(\frac{\mathbb{I}}{d_{ABC}}\right)-b
    \right].
\end{equation}
Substituting this back into \eqref{eq:sigma_stationary} gives the HS orthogonal projection onto the affine constraint set,
\begin{equation}
    \Pi_b(X)
    =
    X-\mathcal{B}^{\dagger}(\mathcal{B}\mathcal{B}^{\dagger})^{+}\bigl(\mathcal{B}(X)-b\bigr).
    \label{eq:affine_projector_ABC}
\end{equation}
In particular, the solution to \eqref{eq:abc_projection_form} is
\begin{equation}
    \rho_{ABC}^{(0)}
    =
    \Pi_b\!\left(\frac{\mathbb{I}}{d_{ABC}}\right).
\end{equation}

We now evaluate this projection explicitly. By definition of the adjoint,
\begin{equation}
    \langle \mathcal{B}(X),(Y_{AB},Y_{BC})\rangle_{\oplus}
    =
    \langle X,\mathcal{B}^{\dagger}(Y_{AB},Y_{BC})\rangle,
\end{equation}
and using
\begin{align}
    \tr\!\big[(\tr_C X)Y_{AB}\big]
    =
    \tr\!\big[X(Y_{AB}\otimes \mathbb{I}_C)\big],
    \\
    \tr\!\big[(\tr_A X)Y_{BC}\big]
    =
    \tr\!\big[X(\mathbb{I}_A\otimes Y_{BC})\big],
\end{align}
we obtain
\begin{equation}
    \mathcal{B}^{\dagger}(Y_{AB},Y_{BC})
    =
    Y_{AB}\otimes \mathbb{I}_C
    +
    \mathbb{I}_A\otimes Y_{BC}.
    \label{eq:Bdagger_ABC}
\end{equation}

Rather than inverting $\mathcal{B}\mathcal{B}^{\dagger}$ explicitly, it is more transparent to compute the action of
$\mathcal{B}^{\dagger}(\mathcal{B}\mathcal{B}^{\dagger})^{+}$ on a consistent pair of marginal shifts.
Let $(X_{AB},X_{BC})$ satisfy the compatibility condition
\begin{equation}
    \tr_A X_{AB}=\tr_C X_{BC}\equiv X_B.
    \label{eq:consistent_pair}
\end{equation}
We claim that the unique minimum-HS-norm correction in the full space that realizes these two shifts is
\begin{equation}
    \Delta(X_{AB},X_{BC})
    \equiv
    \frac{1}{d_C}X_{AB}\otimes \mathbb{I}_C
    +
    \frac{1}{d_A}\mathbb{I}_A\otimes X_{BC}
    -
    \frac{1}{d_A d_C}\mathbb{I}_A\otimes X_B\otimes \mathbb{I}_C.
    \label{eq:explicit_correction_ABC}
\end{equation}
Indeed, a direct check gives
\begin{align}
    \tr_C \Delta(X_{AB},X_{BC})
    &=
    \frac{1}{d_C}\tr_C(X_{AB}\otimes \mathbb{I}_C)
    +
    \frac{1}{d_A}\tr_C(\mathbb{I}_A\otimes X_{BC})
    -
    \frac{1}{d_A d_C}\tr_C(\mathbb{I}_A\otimes X_B\otimes \mathbb{I}_C)
    \nonumber\\
    &=
    X_{AB}
    +
    \frac{1}{d_A}\mathbb{I}_A\otimes X_B
    -
    \frac{1}{d_A}\mathbb{I}_A\otimes X_B
    =
    X_{AB},
\end{align}
and similarly
\begin{align}
    \tr_A \Delta(X_{AB},X_{BC})
    &=
    \frac{1}{d_C}X_B\otimes \mathbb{I}_C
    +
    X_{BC}
    -
    \frac{1}{d_C}X_B\otimes \mathbb{I}_C
    =
    X_{BC}.
\end{align}
Thus \eqref{eq:explicit_correction_ABC} satisfies the required marginal conditions. Moreover,
\begin{equation}
    \Delta(X_{AB},X_{BC})
    =
    \mathcal{B}^{\dagger}\!\left(
        \frac{1}{d_C}X_{AB}-\frac{1}{2d_A d_C}\mathbb{I}_A\otimes X_B,
        \frac{1}{d_A}X_{BC}-\frac{1}{2d_A d_C}X_B\otimes \mathbb{I}_C
    \right),
\end{equation}
so $\Delta(X_{AB},X_{BC})\in \mathrm{ran}(\mathcal{B}^{\dagger})=\ker(\mathcal{B})^{\perp}$.
It is therefore orthogonal to all homogeneous feasible directions and hence is the unique HS-minimal correction. In other words,
\begin{equation}
    \mathcal{B}^{\dagger}(\mathcal{B}\mathcal{B}^{\dagger})^{+}(X_{AB},X_{BC})
    =
    \Delta(X_{AB},X_{BC}).
    \label{eq:explicit_pseudoinverse_action}
\end{equation}

We now apply \eqref{eq:explicit_pseudoinverse_action} to the residual of the maximally mixed state,
\begin{equation}
    \tilde{\rho}_{ABC}=\frac{\mathbb{I}_{ABC}}{d_{ABC}}.
\end{equation}
Its constrained residual is
\begin{equation}
    \mathcal{B}(\tilde{\rho}_{ABC})-b
    =
    \left(
        \frac{\mathbb{I}_{AB}}{d_{AB}}-\rho_{AB},
        \frac{\mathbb{I}_{BC}}{d_{BC}}-\rho_{BC}
    \right),
\end{equation}
whose shared $B$-trace is
\begin{equation}
    X_B
    =
    \tr_A\!\left(\frac{\mathbb{I}_{AB}}{d_{AB}}-\rho_{AB}\right)
    =
    \frac{\mathbb{I}_B}{d_B}-\rho_B
    =
    \tr_C\!\left(\frac{\mathbb{I}_{BC}}{d_{BC}}-\rho_{BC}\right),
\end{equation}
where
\begin{equation}
    \rho_B=\tr_A\rho_{AB}=\tr_C\rho_{BC}.
\end{equation}
Using \eqref{eq:affine_projector_ABC} and \eqref{eq:explicit_pseudoinverse_action}, we obtain
\begin{align}
    \rho_{ABC}^{(0)}
    &=
    \frac{\mathbb{I}_{ABC}}{d_{ABC}}
    -
    \Delta\!\left(
        \frac{\mathbb{I}_{AB}}{d_{AB}}-\rho_{AB},
        \frac{\mathbb{I}_{BC}}{d_{BC}}-\rho_{BC}
    \right)
    \nonumber\\
    &=
    \frac{\mathbb{I}_{ABC}}{d_{ABC}}
    -
    \frac{1}{d_C}\left(\frac{\mathbb{I}_{AB}}{d_{AB}}-\rho_{AB}\right)\otimes \mathbb{I}_C
    -
    \frac{1}{d_A}\mathbb{I}_A\otimes\left(\frac{\mathbb{I}_{BC}}{d_{BC}}-\rho_{BC}\right)
    +
    \frac{1}{d_A d_C}\mathbb{I}_A\otimes\left(\frac{\mathbb{I}_B}{d_B}-\rho_B\right)\otimes \mathbb{I}_C.
\end{align}
The identity contributions cancel exactly, leaving the closed form
\begin{equation}
    \rho_{ABC}^{(0)}
    =
    \rho_{AB}\otimes \frac{\mathbb{I}_C}{d_C}
    +
    \frac{\mathbb{I}_A}{d_A}\otimes \rho_{BC}
    -
    \frac{\mathbb{I}_A}{d_A}\otimes \rho_B\otimes \frac{\mathbb{I}_C}{d_C}.
    \label{eq:zeroth_order_recovery_closed}
\end{equation}
A final direct check confirms that
\begin{equation}
    \tr_C \rho_{ABC}^{(0)}=\rho_{AB},
    \qquad
    \tr_A \rho_{ABC}^{(0)}=\rho_{BC}.
\end{equation}
Therefore \eqref{eq:zeroth_order_recovery_closed} is the unique Hilbert--Schmidt orthogonal projection of the maximally mixed state $\mathbb{I}_{ABC}/d_{ABC}$ onto the affine set defined by the marginal constraints \eqref{eq:marginal_constraints_ABC}. Equivalently, it is the unique minimum-purity representative in this affine Hermitian space, which is the zeroth-order recovery formula used in \rlite{}.

Equation~\xeqref{eq:zeroth_order_recovery_closed} is the unique Hilbert--Schmidt orthogonal projection of the maximally mixed state onto the affine set defined by the marginal constraints.
As in projection-based recovery schemes more generally, this affine projection is not guaranteed to preserve positive semidefiniteness exactly. In practice, positivity can therefore be monitored and, if necessary, restored by a separate repair step. This issue is not specific to the \renyi{} reformulation, but reflects the fact that the recovery problem is solved here at the level of an affine HS projection rather than by an explicit optimization considering the positive-semidefinite condition.

\section{Algorithmic Outline of \rlite{}}
\label{app:pseudocode}

This appendix provides pseudocode for the \rlite{} workflow used throughout the paper.
The aim is to make the sequence of update steps, truncation operations, and bookkeeping explicit at the level of data flow.
Compared to the original \textsc{LITE} pseudocode in Ref.~\cite{artiacoEfficientLargeScaleManyBody2024}, our \rlite{} implementation does not require eigendecompositions or matrix logarithms of the reduced density matrices during time evolution.
Consequently, we can omit the density-matrix ``shifting'' regularization used to stabilize spectral decompositions, and we also do not require additional damping or tolerance parameters in the information removal routine.
In practice this reduces the number of hyperparameters and yields a more stable evolution.
The only additional bookkeeping relative to the non-negative von~Neumann information lattice is the signed R\'enyi-2 split described in Sec.~\ref{sec:renyi2}: the signed purity gain \(\gamma_n^\ell\) is stored for diagnostics, the positive part is used for control thresholds, and the negative part is monitored separately.
\begin{table*}[t]
  \centering
  \caption{Parameters for the \rlite{} workflow used in this work (adapted from the conventions of Ref.~\cite{artiacoEfficientLargeScaleManyBody2024}). Parameters related to eigendecomposition-regularization and damped minimization are not required here and therefore omitted.}
  \label{tab:rlite_params}
  \begin{tabular}{@{}cccc@{}}
    \toprule
    Parameter & Description & Prescription & Numerical value (this work) \\
    \midrule
    $\ell_{\min}$ & Minimization level & Typically $\ell_{\min}\leq\ell_{\max}-1$ & $\ell_{\max}-1$ \\
    $\ell_{\max}$ & Maximum level used for evolution & As large as feasible & $9$ -- $11$ \\
    $q_{\max}$ & Positive-information threshold to activate minimization & Convergence check & $0.5\%$--$2\%$ \\
    $q_{\lstar}$ & Positive-information threshold to increment $\lstar\!\to\!\lstar+1$ & As small as feasible & $\sim 10^{-10}$ \\
    $p$ & Threshold to remove infinite-temperature boundary sites & As small as feasible & $\sim 10^{-7}$ \\
    $r$ & $\lstar$ increment & Hamiltonian locality & $1$ \\
    $\varepsilon$ & Runge-Kutta tolerance & As small as stability allows & $10^{-7}$ \\
    \bottomrule
  \end{tabular}
\end{table*}

Table~\ref{tab:rlite_params} summarizes the parameters and the values used in this work.
Unless stated otherwise, we use $r=1$ and typically choose $\ell_{\min}=\ell_{\max}-1$, such that the active hierarchy level $\lstar$ toggles between the two topmost levels.

\begin{algorithmic}[1]
  \Require Initial reduced density matrices $\{ \rho_n^{\ell_{\min}} \}$, thresholds $q_{\max}$ and $q_{\lstar}$, padding threshold $p$, time step $\delta t$, time integration tolerance $\varepsilon$, final time $t_{\mathrm{final}}$, and levels $\ell_{\min},\ell_{\max}$ (with $r=1$)
  \Ensure  Evolved state $\{ \rho_n^{\ell_{\min}} \}$

  \State $\lstar \gets \ell_{\min}$
  \State $t \gets 0$

  \While{$t < t_{\mathrm{final}}$}

    \State $\{ \rho_n^{\lstar} \} \gets \textproc{Pad-Edges}(\{ \rho_n^{\lstar} \}, p)$
    \Comment Add or remove infinite-temperature boundary sites if needed.

    \State $\{ \rho_n^{\lstar+1} \} \gets \textproc{Petz-Recovery}(\{ \rho_n^{\lstar} \})$
    \Comment Reconstruct higher-level density matrices using the affine projected Petz recovery map.

    \State \textproc{Check-Physicality}($\{ \rho_n^{\lstar} \},\{ \rho_n^{\lstar+1} \}$)
    \Comment Monitor Hermiticity and trace normalization

    \State $\{ \rho_n^{\lstar} \},\delta t' \gets \textproc{Time-Integration}(\{ \rho_n^{\lstar} \}, \{ \rho_n^{\lstar+1} \}, \delta t, \varepsilon)$
    \Comment Propagate the collection of reduced density matrices by one time step and adjust the time step.

    \State \textproc{Compute-Observables}($\{ \rho_n^{\lstar} \}$)
    \Comment Compute diffusion coefficients (e.g.\ spin and energy) from the current hierarchy state

    \State $\{ \gamma_{n,\mathrm{signed}}^{\lstar} \} \gets \textproc{Compute-Signed-Purity-Gain}(\{ \rho_n^{\lstar} \})$
    \Comment Compute $\gamma_n^{\lstar}=\log[(P_{ABC}P_B)/(P_{AB}P_{BC})]$ from purity ratios.

    \State $\gamma_{n,\mathrm{pos}}^{\lstar} \gets \max(\gamma_{n,\mathrm{signed}}^{\lstar},0)$ for all $n$
    \State $\gamma_{n,\mathrm{neg}}^{\lstar} \gets \max(-\gamma_{n,\mathrm{signed}}^{\lstar},0)$ for all $n$
    \Comment Use signed values for diagnostics, positive values for control, and negative values as separate diagnostics.

    \State $\gamma_{\mathrm{pos}}^{\lstar} \gets \sum_n \gamma_{n,\mathrm{pos}}^{\lstar}$
    \State $\gamma_{\mathrm{pos,tot}} \gets \sum_{\ell,n} \max(\gamma_{n,\mathrm{signed}}^\ell,0)$
    \State $\gamma_{\mathrm{neg}}^{\lstar} \gets \sum_n \gamma_{n,\mathrm{neg}}^{\lstar}$
    \Comment Total positive information normalizes threshold tests; negative information is logged but never cancels positive buildup.

    \If{$(\exists n:\; \gamma_{n,\mathrm{pos}}^{\lstar} > q_{\lstar})$ \textbf{and} $\lstar \neq \ell_{\max}$}
      \State $\{ \rho_n^{\lstar+1} \} \gets \textproc{Petz-Recovery}(\{ \rho_n^{\lstar} \})$
      \Comment Increment subsystem hierarchy by one level ($r=1$)
      \State $\lstar \gets \lstar + 1$
      \State $\{ \rho_n^{\lstar} \} \gets \{ \rho_n^{\lstar+1} \}$
    \ElsIf{$\lstar = \ell_{\max}$ \textbf{and} $\dfrac{\gamma_{\mathrm{pos}}^{\lstar}}{\gamma_{\mathrm{pos,tot}}} > q_{\max}$}
      \State $\{ \rho_n^{\ell_{\min}} \} \gets \textproc{Reduce-To-}\ell_{\min}(\{ \rho_n^{\ell_{\max}} \})$
      \Comment Project the working level reduced density matrices back to $\ell_{\min}$ level via partial trace.
      \State $\{ \rho_n^{\ell_{\min}} \} \gets \textproc{Minimize-Info}(\{ \rho_n^{\ell_{\min}} \}, q_{\max})$
      \Comment Remove excess positive local information subject to marginal and current constraints.
      \State \textproc{Check-Physicality-And-Marginals}($\{ \rho_n^{\ell_{\min}} \}$)
      \Comment Recheck after minimization because affine projections/repairs need not preserve positivity and marginals simultaneously.
      \State $\lstar \gets \ell_{\min}$
    \EndIf

    \State $t \gets t + \delta t$
    \State $\delta t \gets \delta t'$
  \EndWhile

  \State \Return $\{ \rho_n^{\lstar} \}$ and diagnostics $\{\gamma_{n,\mathrm{signed}}^\ell,\gamma_{n,\mathrm{neg}}^\ell\}$
\end{algorithmic}
The subroutines \textproc{Minimize-Info} and \textproc{Petz-Recovery} are implemented with the \rlite{}-specific minimization and recovery procedures described in Apps.~\ref{app:proofs1} and~\ref{app:proofs2}, respectively. The time integration \textproc{Time-Integration} is performed using Runge-Kutta 45 with the truncation error threshold given in Table~\ref{tab:rlite_params}.


\end{appendix}

\twocolumngrid

\bibliography{bibliography}

\end{document}